\documentclass[pre,epsfig,article,twocolumn]{revtex4-1}
\def\bfphi{\mbox{\boldmath$\phi$}}

\usepackage{latexsym,amsmath,tabularx,amssymb,bm}
\usepackage{graphicx}

\newcommand \bea{\begin{eqnarray}}
\newcommand \eea{\end{eqnarray}}

\def\simge{\mathrel{%
   \rlap{\raise 0.511ex \hbox{$>$}}{\lower 0.511ex \hbox{$\sim$}}}}
\def\simle{\mathrel{
   \rlap{\raise 0.511ex \hbox{$<$}}{\lower 0.511ex \hbox{$\sim$}}}}

\def\simle{\mathrel{
   \rlap{\raise 0.511ex \hbox{$<$}}{\lower 0.511ex \hbox{$\sim$}}}}

\def\simge{\mathrel{%
    \rlap{\raise 0.511ex \hbox{$>$}}{\lower 0.511ex \hbox{$\sim$}}}}
\def\simle{\mathrel{
    \rlap{\raise 0.511ex \hbox{$<$}}{\lower 0.511ex \hbox{$\sim$}}}}
\newcommand{\beq}{\begin{eqnarray}}
\newcommand{\eeq}{\end{eqnarray}}
\newcommand{\del}{\partial}

\begin{document}

%\title{Scaling function for O(N) models within the exact renormalization group}
\title{Non-perturbative renormalization group preserving full-momentum dependence:  
implementation and quantitative evaluation}

\author{F. Benitez}
\affiliation{LPTMC, CNRS-UMR 7600, Universit\'e Pierre et Marie Curie, 75252 Paris, France}
\affiliation{Instituto de F\`{\i}sica, Facultad de Ciencias, Universidad de la Rep\'ublica, 11400 Montevideo, Uruguay}

\author{J.-P. Blaizot }
\affiliation{Institut de Physique Th\'eorique, CEA-Saclay, 91191 Gif-sur-Yvette, France}

\author{H. Chat\'e}
\affiliation{Service de Physique de l'Etat Condens\'e, CEA-Saclay, 91191 Gif-sur-Yvette, France}

\author{B. Delamotte}
\affiliation{LPTMC, CNRS-UMR 7600, Universit\'e Pierre et Marie Curie, 75252 Paris, France}

\author{R. M\'endez-Galain}
\affiliation{Instituto de F\`{\i}sica, Facultad de Ingenier\'{\i}a, 
Universidad de la Rep\'ublica, 11000 Montevideo, Uruguay}

\author{N. Wschebor}
\affiliation{Instituto de F\`{\i}sica, Facultad de Ingenier\'{\i}a, 
Universidad de la Rep\'ublica, 11000 Montevideo, Uruguay}

\begin{abstract}
We present in detail the implementation of the Blaizot-M\'endez-Wschebor  (BMW) approximation scheme 
of the nonperturbative renormalization group, which allows for the computation of 
the full momentum dependence
of correlation functions. We discuss its signification and its relation with other 
schemes, in particular
the derivative expansion.  Quantitative  results are presented for the
testground of scalar $O(N)$ theories. Besides critical exponents which
are zero-momentum  quantities, we compute in three dimensions in the  whole momentum range
the  two-point function at  criticality and,  in the  high temperature
phase, the  universal structure  factor.  In all  cases, we  find very
good agreement with the best existing results.
\end{abstract}

 % end of PACS codes

\date{\today}% It is always \today, today,
\maketitle

\section{Introduction}
\label{intro}

The  exact  or  non  perturbative  renormalization  group  (NPRG),  as
formulated in the seminal work of Wetterich \cite{Wetterich93}, leads
to an  exact flow equation  for an effective action 
(see also Refs. \cite{Ellwanger93,Wilson} for the original formulation of the NPRG).  This equation
cannot be solved in general, but offers the possibility of developing
approximation schemes qualitatively different from those based on
perturbation theory, allowing us in particular to tackle non-perturbative problems.

The  ``derivative expansion''  (DE)  is  such a  scheme:  based on  an
expansion of the running effective action in terms of gradients of the
fields,  it has  been applied  successfully to  a variety  of physical
problems,  in  condensed  matter,  particle physics or statistical mechanics  (see
e.g.  \cite{Berges02}). The DE
scheme  allows  us  to  calculate not only universal,  but  also  non-universal
quantities defined  at vanishing  momenta, such as  critical exponents
and phase  diagrams (see  e.g. \cite{Berges02,delamotte03,ref6}). However, it does
not  give access to  the full  momentum dependence  of correlation
functions, something desirable in many situations.

The present paper deals with
another scheme, namely with the strategy  proposed by Blaizot, M\'endez-Galain  and Wschebor (BMW)
in    \cite{Blaizot:2005xy,Blaizot:2005wd}   which   is
reminiscent of earlier attempts by  Parola and
Reatto \cite{Parola}.  The BMW scheme makes use of the cutoff function
$R_k(q)$, typical of NPRG  studies, which renders all vertex functions
smooth, and insures that the flow at scale $k$ involves the integration
of fluctuations with momenta $q$ {\it at most} of order $k$.  At order
$s$, the  approximation consists in setting the  internal momentum $q$
to zero  in the vertex functions  of order larger than  $s$, leaving a
closed set  of flow equations for the  first $s$ vertex  functions. It
was  shown in \cite{Blaizot:2005xy} that the BMW  method encompasses
any perturbative results, provided it  is pushed to
high-enough  orders $s$. For the particular $s=2$ case studied in the following it is 
 one-loop exact for the two-point function. It has also been shown in 
\cite{Blaizot:2005xy} that in the   $N\to\infty$ limit of the $O(N)$ scalar models 
the BMW scheme becomes exact and allows for the computation of
 any correlation functions.

The BMW  approximation scheme has  been applied to $O(N)$  models with
success, in simplified  versions  involving either  expansions in  the
fields  \cite{Guerra}  or  an approximated  propagator  \cite{BMWnum}.
In this paper, we provide a full account
of  the  practical  implementation  of the  method without further 
simplifications. We also detail and extend the
first results obtained recently at order $s=2$ \cite{Benitez09}.
In particular, we extract the universal scaling function governing
the critical region. At the quantitative level, we find very good agreement with 
the best existing results. 

The remainder of  this paper is organized as  follows: after a general
presentation of  the NPRG  framework (Sect.\ref{NPRG}), we  detail the
BMW  approximation scheme  in  Section~\ref{method}, and  show how  to
implement    it   in    practice    in   Section~\ref{num_impl}.    In
Section~\ref{criticality},  we  report  on  the  results  obtained  at
criticality  (critical exponent values,  shape of  two-point function,
etc.), while in Section~\ref{scaling} we compare the universal scaling
function obtained  in the whole  critical region to  existing results.
In Section~\ref{sectionDE}, we show how the BMW results shed new light
on the  derivative expansion approximation  scheme, and we  discuss in
particular  its   validity  domain.   Conclusions  can  be   found  in
Section~\ref{Conclusions},  while   the  appendices  gather  technical
material.

\section{The NPRG framework}
\label{NPRG}

For  the  sake of  simplicity,  in the  following,  we only
discuss a scalar field theory in  the Ising universality
class, and  defer to Appendix~\ref{ONmodels} the  presentation of the
equations that hold for general $O(N)$-symmetric models.

We consider the  usual partition function
\beq\label{partition}
{\cal Z}[j]=\int {\cal D}\varphi \, {\rm e}^{-S+\int_x j \varphi}
\eeq
with the classical action
\beq\label{eactON} 
S = \int {d}^{d}x\,\left\lbrace{ \frac{1}{2}}   \left(\del_\mu
\varphi\right)^2  + \frac{m^2}{2} \, \varphi^2 + \frac{u}{4!}
\,\varphi^4 \right\rbrace \,.  
\eeq
In Eq.~(\ref{partition}),  $j(x)$ is an external source and  $\int_x j\varphi$ is a
shorthand for $\int d^dx  \,j(x)\varphi(x)$.

The NPRG strategy is to build a family of theories indexed by a 
momentum scale parameter $k$,   
such that fluctuations are smoothly taken into account  as $k$ is lowered 
from the  microscopic scale $\Lambda$ down  to 0
\cite{Wetterich93,Ellwanger93,Tetradis94,Morris94,Morris94c,Bagnuls:2000ae,Berges02,delamotte07}). 
In practice, this is achieved by adding  to the original  
Euclidean action $S$ a $k$-dependent quadratic (mass-like) term of the form   
\beq\Delta S_k[\varphi]= \frac{1}{2} \int_q\: R_k(q)\varphi(q)\varphi(-q), \eeq
with
\beq \int_q\equiv\int \frac{d^dq}{(2\pi)^d}, \notag\eeq 
so that the partition function at scale $k$ reads
\beq
{\cal Z}_k[j]=\int {\cal D}\varphi \, {\rm e}^{-S-\Delta S_k+\int_x j\varphi}\, .
\eeq
The cut-off function $R_k(q)$ is chosen so that: 
i) it is of order $k^2$  for $q\ll k$, which effectively suppresses the modes $\varphi(q\ll  k)$;  
ii) it vanishes for $q\gg k$, leaving the modes  $\varphi(q)$ with  $q\gg k$ unaffected. 
Thus,  when $k=\Lambda$, $R_k(q)$ is of order $\Lambda^2$ for all  $q\le\Lambda$,  
and  fluctuations are essentially frozen. On the other hand, when $k=0$, $R_k(q)$  vanishes identically 
so that ${\cal Z}_{k=0}={\cal Z}$, and the original theory is recovered. The specific form of 
the cut-off function $R_k(q)$ will be specified later. 

Following Wetterich \cite{Wetterich93}, an effective action at scale $k$,  
$\Gamma_k[\phi]$, is defined through the (slightly modified) Legendre transform
\beq
\Gamma_k[\phi]+\log {\cal Z}_k[j]=\int_x j\phi -\Delta S_k[\phi], 
\eeq
where $\phi(x)=\delta \ln {\cal Z}_k[j]/\delta j(x)$.
This effective action obeys the exact flow equation \cite{Wetterich93} (up to a volume factor):
\beq \label{NPRGeq}
\partial_t \Gamma_k[\phi]=\frac{1}{2} \int_q\,\frac{\partial_t R_k(q)}{\Gamma_k^{(2)}[q,-q;\phi]+R_k(q)}
\eeq
where $ \partial_t\equiv k \partial_k$ and  $\Gamma_k^{(2)}[q,-q;\phi]$ is the Fourier transform
of  the second functional derivative of $\Gamma_k[\phi]$:
\beq
\Gamma_k^{(2)}[x_1,x_2;\phi]\equiv  \frac{\delta^2\Gamma_k}{\delta\phi(x_1)
\delta\phi(x_2)} \;.
\eeq
Thus, $(\Gamma_k^{(2)}[q,-q;\phi]+R_k(q))^{-1}$ 
is the full propagator in the presence of the field $\phi(x)$.
The initial condition of the flow equation (\ref{NPRGeq}) is specified at 
the microscopic scale $k=\Lambda$ where fluctuations are frozen by  
$\Delta S_k$, so that $\Gamma_{k=\Lambda}[\phi]\approx S[\phi]$. 
The effective action $\Gamma[\phi]$ of the original scalar field theory  is obtained 
as the solution of  Eq.~(\ref{NPRGeq}) for $k\to 0$, at which point  $R_k(q)$ 
vanishes identically. 

When $\phi$ is constant, the functional $\Gamma_k[\phi]$ reduces, to within a volume factor $\Omega$, to the 
effective potential $V_k(\phi)$:
\beq
\Gamma_k[\phi]=\Omega V_k(\phi),\qquad \phi {\mbox  \; { \rm constant.}}
\eeq
The  flow equation for   $V_k$ reads  
\beq\label{eqforV} 
\partial_t
V_k(\rho)=\frac{1}{2}\int _q \,\partial_t R_k(q)\, G_k(q,\phi),
\eeq
where
\begin{equation}\label{G-gamma2}
G^{-1}_{k} (q,\phi) =\Gamma^{(2)}_{k} (q,\phi) + R_k(q),
\end{equation}
(see Appendix \ref{Notations} for notation).

Let us now consider the flow of  $n$-point functions.
By taking two functional derivatives of Eq.~(\ref{NPRGeq}), letting $\phi$ be constant, and Fourier transforming, 
one obtains the equation for the 2-point function:
\begin{widetext}
\begin{eqnarray}
\label{gamma2champnonnul}
\partial_t\Gamma_{k}^{(2)}(p,\phi)=\int_q
\partial_t R_k(q) G_{k}^2(q,\phi)\bigg\{&&\hspace{-4mm}\Gamma_{k}^{(3)}(p,q,-p-q,\phi) G_{k}(q+p,\phi)\Gamma_{k}^{(3)}(-p,p+q,-q,\phi)\nonumber\\
&&\hspace{-4mm} -\frac{1}{2}\Gamma_{k}^{(4)}
(p,-p,q,-q,\phi)\bigg\}.
\label{exact-flow-gamma2}
\end{eqnarray}
\end{widetext}

The flow equations (\ref{eqforV}) and (\ref{gamma2champnonnul}) are the first equations of an infinite 
tower of coupled equations for the $n$-point functions: typically the equation for $\Gamma_k^{(n)}$ 
involves all the vertex functions up to $\Gamma_k^{(n+2)}$.   
Approximations and truncations are thus needed to obtain any practical result.

The presence  of a  sufficiently-smooth cutoff function  $R_k(q)$, (i)
insures that  the $\Gamma^{(n)}_k$'s  remain regular functions  of the
momenta  and (ii) limits, through the term $\partial_t R_k(q)$, 
the internal momentum $q$ in equations such as
Eq.~(\ref{gamma2champnonnul}),  to $q\lesssim  k$.   These key  remarks
allow  for  approximations  without  equivalent  in  more  traditional
frameworks and  thus constitute one  of the specificities of  the NPRG
approach.

The approximation  scheme most  widely used  is  the derivative
expansion,  which is  entirely based  on the  above remarks  about the
analyticity of the vertex functions.  It  amounts to formulating an ansatz
for $\Gamma_k[\phi]$ as an expansion in the derivatives
of the field. For instance, at order $\nabla^2$:
\begin{equation}  
\Gamma_k[\phi]=\int_x\left(  V_k(\phi)  +\frac{1}{2}
Z_k(\phi)\left( \nabla\phi \right)^2 +\mathcal{O}(\nabla^4) \right).
\label{ordre2}
\end{equation} 
The flow equation  (\ref{NPRGeq}) then reduces to a set
of  two  coupled, partial  differential  equations  for the  functions
$V_k(\phi)$  and  $Z_k(\phi)$.  The  derivative  expansion scheme  has
produced,   along  the   years,   a  wealth   of  remarkable   results
(see e.g. \cite{Berges02,delamotte03,tarjus04})   but  it does  not
allow to access the  full-momentum dependence of the vertex functions,
something inherently  possible within the BMW scheme.  We  discuss this point
in Section~\ref{sectionDE}, where we show how the BMW approach sheds  new light
 on the derivative expansion.

\section{The BMW approximation scheme}
\label{method} 

The  BMW scheme  at order  $s$ aims  at preserving  the  full momentum
dependence   of  $\Gamma^{(s)}_k$   and  approximating   the  momentum
dependence of  $\Gamma^{(s+1)}_k$ and $ \Gamma^{(s+2)}_k$  in the flow
equation of $\Gamma^{(s)}_k$ \cite{Blaizot:2005xy,Blaizot:2005wd}.

For uniform fields, the following formula:
\beq\label{Gamman0}
\Gamma^{(s+1)}_k(\left\{p_i\right\},0,\phi)=
\partial_\phi \Gamma^{(s)}_k(\left\{p_i\right\},\phi)
\eeq
where the index $i$ runs between 1 and $s$, allows to reduce the order of 
the vertex functions as soon as one momentum is vanishing. 

The  BMW  approximation relies  on  this  formula,  together with the
analyticity of  the vertex  functions and the  fact that  the internal
momentum     $q$     in     the     flow     equations     such     as
Eq.~(\ref{gamma2champnonnul})  is  effectively  limited to  $q\lesssim
k$. The BMW scheme at order  $s$ thus consists in:

(i) neglecting the dependence on the internal momentum $q$ of $\Gamma^{(s+1)}_k$ 
and $ \Gamma^{(s+2)}_k$:
\begin{multline}
\label{BMWapprox}
\Gamma^{(s+1)}_k(p_1,\dots,p_s-q,q,\phi) \to  \\ 
\Gamma^{(s+1)}_k(p_1,\dots,p_s,0,\phi),
\end{multline}
and similarly for $\Gamma^{(s+2)}_k(p_1,\dots,p_s,-q,q,\phi)$;

(ii) using Eq.~(\ref{Gamman0}) which allows us to express the approximated expressions 
(\ref{BMWapprox}) as derivatives of 
$\Gamma^{(s)}_k$ with respect to $\phi$, thereby closing the hierarchy of RG equations at the level of 
the flow equation for $\Gamma^{(s)}_k$. 

Note that  the substitution in Eq.(\ref{BMWapprox}) is {\it not}  applied
  to  the  $q$-dependence already  present  in the  bare
$n$-point  functions  \cite{Blaizot:2005xy,Blaizot:2005wd}. Thus,  for
instance, at the  lowest level of the approximation  ($s=0$, the local
potential  approximation  discussed  below  in  Sect.~\ref{LPA}),  one
leaves untouched the bare $q^2$ dependence of $\Gamma^{(2)}_k(q)$. This ensures
in particular that the propagator is one-loop exact.

The accuracy of the scheme depends  on the rank $s$ at which
one  operates  the  approximation.  Obviously  the  implementation
becomes increasingly complicated as $s$ grows. We will show later that
good results can be obtained  with low order truncations, i.e., at the
levels $s=0$ and $s=2$. The corresponding approximations are discussed
in the next subsections.

 \subsection{$s=0$: The local potential approximation}
\label{LPA}

The local potential  approximation (LPA) is often seen  as the leading
order   of  the  DE   approximation  scheme
\cite{Berges02,delamotte07}. In  this subsection, we show  that it can
be seen also as the zeroth order of the BMW scheme.

The BMW approximation for $s=0$, consists in neglecting the (nontrivial)
$q$-dependence of the 2-point function
in the flow equation (\ref{eqforV}) of the ``zero-point'' function, that 
is of the effective potential $V_k$.
That is, one substitutes
\beq\label{GLPA}
\Gamma^{(2)}_k (q,\phi)\to q^2+\Gamma^{(2)}_k (0,\phi)=q^2+\partial_\phi^2 V_k\,.
\label{approx_LPA}
\eeq
Note that the  equality in the equation above is a particular case of the general relation (\ref{Gamman0}).
By  substituting  Eq.(\ref{approx_LPA}) in Eq.(\ref{eqforV}), one gets the equation for the potential in the form
\beq\label{VLPA}
\partial_t V_k(\rho)=\frac{1}{2}\int_q \; \frac{\partial_t R_k(q)}{q^2+R_k(q)+\partial_\phi^2 V_k}.
\eeq
This is the flow equation for the potential obtained within the DE  truncated at 
the LPA level. 
There, Eq.(\ref{eqforV}) is derived by computing the propagator from the ansatz
\beq\label{GammaLPA}
\Gamma_k^{\rm LPA}[\phi]=\int d^dx\left\{ \frac{1}{2} \left( \nabla\phi\right)^2 +V_k(\phi)\right\}
\eeq
and inserting it  Eq.(\ref{eqforV}).

Since it allows for the calculation of the entire effective potential, the  LPA  provides
information on  {\em all} the $\Gamma^{(n)}_k$'s at once but only for vanishing external 
momenta: these functions are indeed those that are obtained by taking the derivatives of the 
effective potential, i.e., 
\beq\label{Vgenerating}
\Gamma_k^{(n)}(0,\cdots , 0,\phi)=\del_{\phi}^n V_k.
\eeq
Non trivial momentum dependence will appear at the next level of approximation, to be 
described in the next subsection. 

\subsection{First order with full momentum dependence: $s=2$}

The order $s=2$ is the first order of the approximation where a non-trivial momentum dependence is kept.
The loop momentum $q$ in the 3 and 4-point functions 
in the right hand side of Eq.~(\ref{gamma2champnonnul}) is neglected, and 
  Eq.~(\ref{Gamman0}) is applied. The flow equation for $\Gamma_k^{(2)}(p,\phi )$ becomes then a closed equation
 \begin{equation}
 \label{2pointcloseda}
 \partial_t\Gamma_k^{(2)}= J_3(p,\phi ) \left( \partial_\phi \Gamma_k^{(2)} \right)^2
\! -\frac{1}{2} I_2(\phi) \partial^2_\phi \Gamma_k^{(2)},
\end{equation}
where we have introduced the notation:
\begin{multline} \label{defJ}
I_n(\phi) \equiv J_n(p=0,\phi),\\ J_n(p,\phi )\equiv \int_q \;\partial_t R_k(q)\; G_k(p+q,\phi )G^{n-1}_k(q,\phi). 
\end{multline} 

Again, as was the case for the LPA, the approximation at $s=2$ provides 
information on {\em all} the $n$-point functions. 
This time, the $n$-point functions depend on a single momentum. They may be 
obtained as derivatives of the 2-point function, 
according to 
\beq\label{gammagener}
\Gamma_k^{(n)}(p,-p,0,\cdots , 0,\phi)=\del_{\phi}^{n-2} \Gamma_{k}^{(2)}(p,\phi),
\eeq
which may be viewed as a generalization of  Eq.~(\ref{Vgenerating}). 
Thus, for instance, the momentum dependence that remains within the 
3 and 4-point vertices in  Eq.~(\ref{2pointcloseda}) is 
indeed that of the 2-point function itself.   

At this point an important subtlety appears, coming from the fact that the flow of the
potential (or of its second derivative) can be calculated either from Eq.(\ref{eqforV}), 
in which $G_k(q,\phi)$ is obtained from
(\ref{2pointcloseda}) and (\ref{G-gamma2}),
or directly from Eq.(\ref{2pointcloseda}) at $p=0$, since  $\Gamma_{k}^{(2)}(0,\phi)=\partial_\phi^2 V_k$.
If no approximations were
made, both results would be identical. However, once approximations are done, as it is the case here, both results
do not coincide. 

At any order $s$ of the BMW approximation scheme, the same ambiguity takes place for
any correlation function $\Gamma_k^{(n)}$ up to $n=s-1$. Given the fact that the approximation is imposed
only on the flow equation of $\Gamma_k^{(s)}$ and not on those of the $\Gamma_k^{(n)}$ with $n<s$, 
it is natural to compute these
functions from their own flow equation (which is exact) and not from the flow equation of $\Gamma_k^{(s)}$ 
(which is approximate). 

One then  subtracts from $\Gamma_k^{(s)}$ the parts of it that can be expressed in terms of lower order
correlation functions, and perform the BMW approximation in the equation for the {\it difference}.
For $s=2$, this amounts to computing the potential from Eq.(\ref{eqforV}) and to implementing the BMW approximation
on  $\Gamma_{k}^{(2)}(p,\phi)-\Gamma_{k}^{(2)}(0,\phi)$.

The rationale behind this choice is that computing the flow of $\del_\phi^2 V_k$ from the equation for
$\Gamma^{(2)}_k (p,\phi)$ at $p=0$ would imply two approximations: the equation
(\ref{2pointcloseda}) for
$\Gamma^{(2)}_k (p,\phi)$ is itself approximated, and propagators and vertices in its r.h.s. are
also approximated. On the contrary, Eq.(\ref{eqforV}) for the potential is formally exact and only the
propagator used in it is approximated. This general consideration can be made more concrete in the perturbative
regime: The function $\Gamma^{(2)}_k (p,\phi)$ obtained from Eq.(\ref{2pointcloseda})
is one-loop exact and so is $\Gamma^{(2)}_k (p=0,\phi)=\del_\phi^2 V_k$. When  the
corresponding propagator, computed from Eq.(\ref{G-gamma2}), is inserted in Eq.(\ref{eqforV}), 
the obtained potential becomes two-loop exact.
By generalizing the above subtraction procedure at higher orders similar perturbative 
considerations can be made: at order $s=2s'$ of the
BMW scheme, the potential computed from Eqs.(\ref{eqforV},\ref{G-gamma2}) is 
$(s'+1)-$loop exact, $\Gamma^{(2)}_k$ computed from Eq.(\ref{exact-flow-gamma2}) is $s'-$loop exact, and so on.
We thus expect
that implementing the BMW approximation only on  the part of  $\Gamma_k^{(s)}$ which is genuinely of order $s$,
 will have a decreasing impact on the lower order correlation functions as $s$ grows.

In practice, for $s=2$, we rewrite
\beq\label{defsigmabis}
\Gamma^{(2)}_k (p,\phi)=  p^2+ \Delta_k (p,\phi)+ \partial_\phi^2 V_k(\phi)  ,
\eeq
where $V_k(\phi)$ is obtained by solving  (\ref{eqforV}), and moreover,
for numerical convenience (see below), the bare $p^2$ term has been  extracted.
The BMW approximation is implemented only on the flow for $\Delta_k$.
The equation for $\Delta_k(p,\phi)$ can be deduced from (\ref{2pointcloseda}) by subtracting its $p=0$ form:
\begin{widetext}
\begin{multline}\label{2pointclosedab0}
\partial_t\Delta_k(p,\rho)=
2\rho  J_3(p,\rho) \; \left[u_k(\rho)+\Delta_k^\prime(p,\rho)\right]^2-2\rho I_3(\rho)\; u_k^2(\rho)
-\frac{1}{2} I_2(\rho) \; \left[\Delta_k^\prime(p,\rho)+2\rho\Delta_k^{\prime\prime}(p,\rho)\right],
\end{multline}
\end{widetext}
with
\begin{align}
&\rho=\frac{1}{2}\phi^2\\
&m_k^2(\rho)\equiv\Gamma_k^{(2)}(0,\rho)=\del_\phi^2V_k\\
&u_k(\rho)\equiv  \partial_\rho m_k^2(\rho) \;,
\end{align}
and the symbol $^\prime$ denotes the derivative with respect to $\rho$.

In closing this section, let us mention that the relationships between the BMW scheme at order  $s=2$
and, on one the hand the large $N$ expansion and, on the other hand the DE, are discussed 
respectively in Sections~\ref{critical-exponents} and \ref{sectionDE}.

%%%%%%%%%%%%%%%%%%%%%%%%

\section{Implementation at criticality}
\label{num_impl}

In order to treat efficiently the low momentum region at criticality and, in particular, 
to capture accurately the fixed point structure, we first introduce dimensionless and 
renormalized variables, to be denoted
with a tilda.
We thus introduce a renormalization factor $Z_k$, which reflects the finite change of 
normalization of the field between the ultraviolet scale $\Lambda$ and the scale $k$.
Within the DE at $O(\nabla^2)$ this factor describes the overall variation with $k$ 
of the function $Z_k(\phi)$ in Eq.(\ref{ordre2}). We define here $Z_k$ by
\begin{equation}\label{Zkapparho}
Z_k=\left.\frac{\partial \Gamma^{(2)}_{k}(p,\rho)}{\partial p^2}\right|_{p=p_0; \rho=\rho_0},
\end{equation}
where $p_0$ and $\rho_0$ are a priori arbitrary. From $Z_k$ we define the running anomalous dimension $\eta_k$ by
\beq
\eta_k= -k\del_k \ln Z_k.
\label{etak}
\eeq

Momenta are naturally rescaled according to $p=k \tilde p$.
Other quantities are made dimensionless by dividing 
them by appropriate powers of $k$ (and possibly conveniently extracting numerical factors). Thus we define
\begin{align}
 &\rho =K_dk^{d-2}Z_k^{-1}\tilde\rho,\quad 
m^2_k(\rho)= Z_kk^2\, \tilde m_k^2(\tilde\rho), \nonumber \\ 
& u_k(\rho) =Z_k^2 k^{4-d}K_d^{-1} \,\tilde u_k(\tilde\rho),
\label{dimensionless_quantities}
\end{align}
where $K_d$ is a constant originating from angular integrals,  
\beq K_d^{-1}=2^{d-1}d\pi^{d/2}\Gamma(d/2).\eeq 
We also set
 \begin{align}  
G_k(p,\rho)&=\frac{1}{Z_k k^2}\tilde G_k(\tilde
p,\tilde\rho) ,\\ J_n(p,\rho) &=K_d\,
\frac{k^{d+2-2n}}{Z_k^{n-1}}\tilde
J_n(\tilde p,\tilde\rho).
\end{align}
Instead of $ R_k(q)$ it is convenient to work with a dimensionless cut-off function considered 
as a function of $y=q^2/k^2$: 
 \beq
 r(y)\equiv \frac{R_k(q)}{q^2 Z_k}.
\eeq

Now, we note that as $p\to 0$ at fixed $k$, $\Delta_k(p,\rho)\propto p^2$. This  $p^2$-dependence  may generate 
 numerical instabilities in the equation for $\Delta$ (once transformed to dimensionless variables).
In order to avoid these, we found it convenient to introduce the  renormalized
and dimensionless 2-point function $\tilde{Y}_k(\tilde p,\tilde\rho)$:
\begin{eqnarray}
\label{def-Y-tilde}
1+\frac{\Delta_k(p,\rho)}{p^2}
\equiv Z_k (1+\tilde{Y}_k(\tilde p,\tilde\rho)),
\end{eqnarray}
The function $\tilde{Y}_k(\tilde p,\tilde\rho)$ is a slowly varying function of $\tilde p$ and its flow equation is regular. 
This equation is easily obtained from the flow equation for $\Delta(p,\rho)$, Eq.~(\ref{2pointclosedab0}). It reads
\begin{eqnarray}\label{eqnforY}
 \del_t \tilde{Y}_k =&&\hspace{-3mm}\eta_k  (1+\tilde{Y}_k) +\tilde p\,\del_{\tilde p}\tilde{Y}_k-(2-d-\eta_k)\tilde \rho
\tilde{Y}_k^\prime\nonumber\\
&&+2\tilde\rho\, \tilde p\,^{-2}[( \tilde p\,^{2}\tilde{Y}_k'+\tilde u_k)^2 \tilde J_3- \tilde u_k^2\tilde I_3]\nonumber\\
&&-\tilde I_2(\tilde{Y}_k'/2+\tilde\rho\tilde{Y}_k'') \;.
\end{eqnarray}

The normalization condition (\ref{Zkapparho}) is now expressed as:
\beq\label{renorm3}
\tilde{Y}_k(\tilde p=\tilde p_0, \tilde \rho =\tilde\rho_0)=0, \quad \forall k .
\eeq

The equation for $\tilde{Y}_k(\tilde p,\tilde\rho)$ needs to be completed by the flow equation for the dimensionless 
effective potential $\tilde{V}_k(\tilde\rho)=k^{-d} V_k(\rho)$, or rather, the equation for its derivative 
 $\tilde{W}_k(\tilde\rho)=\tilde{V}_k'(\tilde\rho)$, which is more convenient
since  $\tilde{V}_k$ contains a trivial constant part which
induces a numerical divergence. The equation for $\tilde{W}_k$ reads:
\begin{multline}
\partial_t \tilde{W}_k(\tilde \rho)=-(2-\eta_k) \tilde{W}_k(\tilde \rho)
+(d-2+\eta_k)\tilde \rho\,   \tilde{W}_k'(\tilde \rho)\\ 
+\frac{1}{2}   \tilde I_1'(\tilde\rho).\ \ \ \ \ \ \ \ \ \ \ \ \ \ \ \ \ \ 
\end{multline}

Finally $\eta_k$ in Eq.(\ref{eqnforY}) is implicitly determined by inserting the renormalization condition  (\ref{renorm3})
in the flow equation of $\tilde{Y}_k(\tilde p,\tilde\rho)$  and evaluating the r.h.s. at $\tilde \rho =\tilde\rho_0$
and $\tilde p=\tilde p_0$. 

In principle, and if no approximations were performed, no physical quantity would depend on the choice of $\tilde \rho_0$ and $\tilde p_0$,
nor on the relatively free choice of the cut-off function $r(y)$.
In practice, the BMW scheme, as any approximation scheme, introduces  spurious dependence on these choices. 
Below, we use the one-parameter family of cut-off functions:
\beq
r(y)=\frac{\alpha}{{\rm e}^{y}-1} \;,
\label{cut-off-exp}
\eeq
and study the dependence of our results on  $\tilde \rho_0$, $\tilde p_0$, and $\alpha$ \cite{Canet03}.

We numerically solved the flow equations from given initial conditions:
\begin{equation}
\begin{array}{ll}
\tilde W_\Lambda(\tilde\rho)&=r/\Lambda^2+(u/3)\Lambda^{d-4} K_d\tilde \rho\\
\tilde Y_\Lambda(\tilde\rho,\tilde p)&=0 \;,
\end{array}
\label{cond-init}
\end{equation}
searching for the critical point by dichotomy on the initial parameters. We 
 set $\tilde u\equiv u \Lambda^{d-4}K_d$.

The numerical resolution is done on a fixed, regular, $(\tilde{p},\tilde{\rho})$ grid, 
with $0\le \tilde p\le\tilde{p}_{\rm max}$ and 
$0\le \tilde \rho\le \tilde\rho_{max}$. With our choice of cut-off function, the contribution 
of the momentum interval $\tilde{q}\in [4,\infty]$ to the integrals $\tilde{I}_n$ and $\tilde{J}_3$
 is extremely small and thus we neglect it by restricting the integration domain to  $\tilde{q}\in [0,4]$.
When computing the 
double integrals $\tilde{J}_3(\tilde{p},\tilde{\rho})$, we need to evaluate $\tilde{Y}$
for momenta $\tilde{p}+\tilde{q}$  beyond $\tilde{p}_{\rm max}$. 
In such cases, we set $\tilde{Y}(\tilde{p}>\tilde{p}_{\rm max})=\tilde{Y}(\tilde{p}_{\rm max})$, an approximation
checked to be excellent for $\tilde{p}_{\rm max}\ge 5$.
To access the full momentum dependence, 
we also calculate $\Gamma^{(2)}_k(p,\tilde{\rho})$ 
 at a set of fixed, freely chosen, external $p$ values. For a given such $p$, $p/k$ is
within the grid at the beginning of the flow. 
This is no longer so when $k<p/\tilde{p}_{\rm max}$; then, we switch to the 
dimensionful version of Eq.~(\ref{eqnforY}), and  also set 
$J_3(p,\tilde{\rho})=G(p,\tilde{\rho})J_2(0,\tilde{\rho})$, 
an excellent  approximation when $p>k\,\tilde{p}_{\rm max}$.

\begin{figure}[tp]
\begin{center}
\includegraphics[width=\columnwidth,clip]{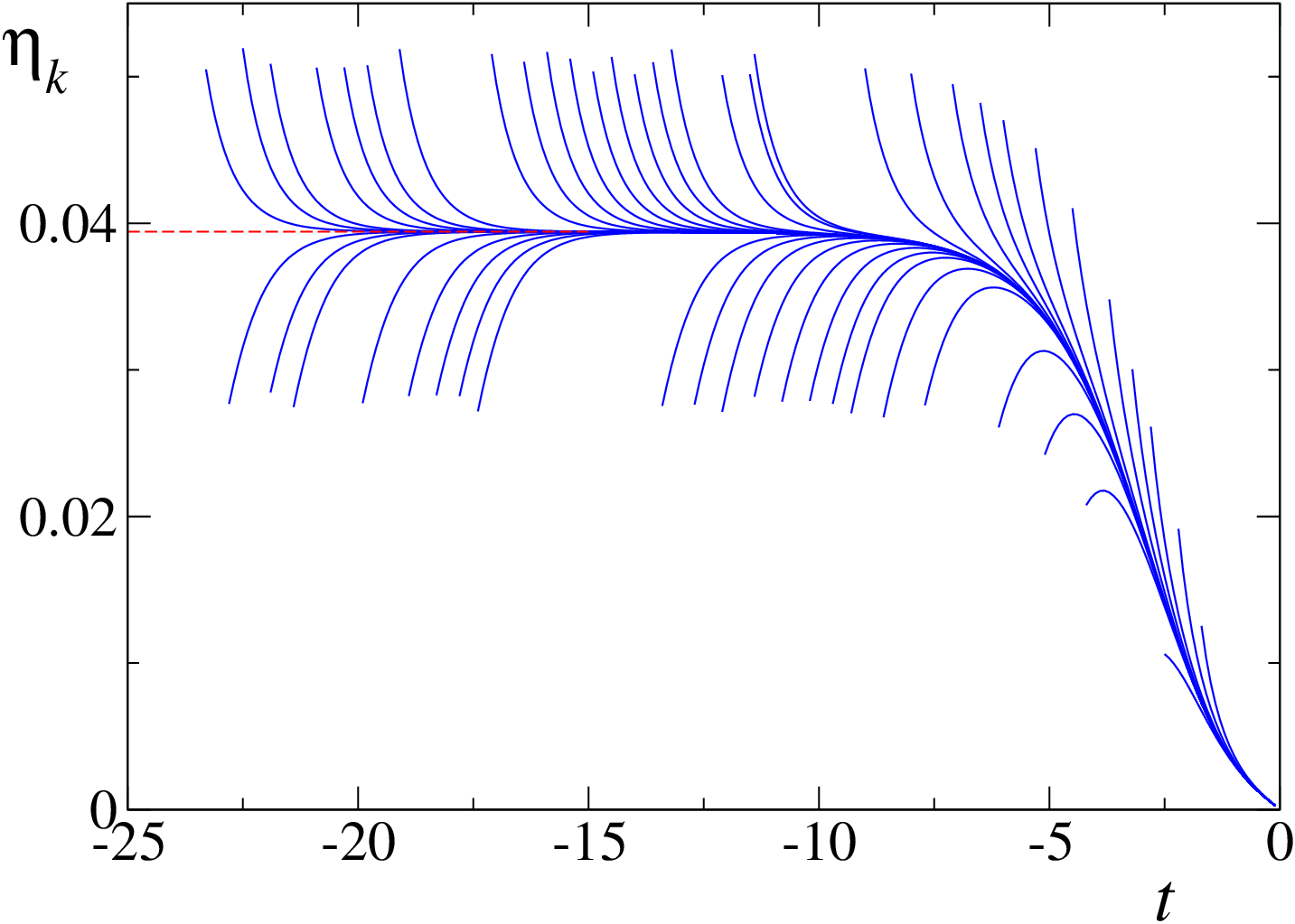}
\end{center}
\caption{\label{etafix}(color online) Example of the dichotomy procedure used for reaching the fixed point 
($N=1$, $d=3$,  $\tilde u=6 \times 10^{-2}, \tilde\rho_0=0, \tilde p_0=0$,  $\alpha=2.25$).
The plot shows the running
anomalous dimension $\eta_k$ as a function of $t=\ln(k/\Lambda)$. 
Each curve corresponds to a different initial value of $r$ (see Eq.(\ref{cond-init})).
The red dashed line indicates the estimated asymptotic value $\eta\simeq 0.03943$.}
\end{figure}

We found that the simplest time-stepping (explicit Euler), a
finite-difference evaluation of derivatives on a regular 
$(\tilde{p},\tilde{\rho})$ grid, and the use of Simpson's rule to 
calculate integrals, are sufficient to produce stable and 
fast-converging results. For  all the 
quantities calculated, the convergence to at least three significant digits  is reached with  
a $(\tilde{p},\tilde{\rho})$ grid of $50\times60$ points and elementary steps
$\delta\tilde{p}=0.1$ and $\delta\tilde{\rho}=0.1$.
With such a grid, a 
typical run takes a few minutes on a current personal computer. The step in 
$t=\log k/\Lambda$ is $\Delta t=10^{-4}$, and the flow is run down to $t\sim -20$. 
In order to find the  fixed point, we performed a simple dichotomy 
procedure on the bare mass  $m^2_\Lambda=r/\Lambda^2$ at fixed $ u$,
by studying the flow of  $\tilde W_k(0)$.
Fig.~\ref{etafix} illustrates the flow of $\eta_k$ as one approaches the fixed point. 

%%%%%%%%%%%%%%%%%%%%%%%%%%%%%%%%%%%%%%%%%%%%%%%%%%%%%%%%%

\section{Results at criticality}
\label{criticality}

Although the main goal of the BMW method is to provide access to full-momentum dependence,  
it can of course also be used  to compute critical 
exponents and other zero-momentum quantities \cite{Benitez09}.
 In this section, where we return to the $O(N)$ models
(with general $N$), we provide details on the calculation 
of the critical  exponents and check their robustness with respect to variations of the 
different parameters of the method such as the numerical resolution, 
the choice of the cut-off function and the location of the normalization point 
($\tilde\rho_0$,$\tilde p_0$). 

Since we focus here on the regime of small momenta, 
it is convenient to take as initial condition a  value of the dimensionless
coupling $\tilde u$ not too small compared to 1 in order to initialize the 
flow far from the Gaussian fixed point $\tilde u=0$ and thus to 
approach quickly the infrared fixed point. 
This is useful, not only because of the shortened time 
needed to reach the critical regime, but also
because otherwise, due to the 16-digit precision used, 
our dichotomy procedure does not allow 
for an accurate determination of the fixed point directly from initial
parameters. The results to be presented below have been calculated 
for $\tilde u=6 \times 10^{-2}/N$. 

\subsection{Numerical extraction of critical exponents} 

\begin{figure}[t]
\begin{center}
\includegraphics[width=\columnwidth,clip]{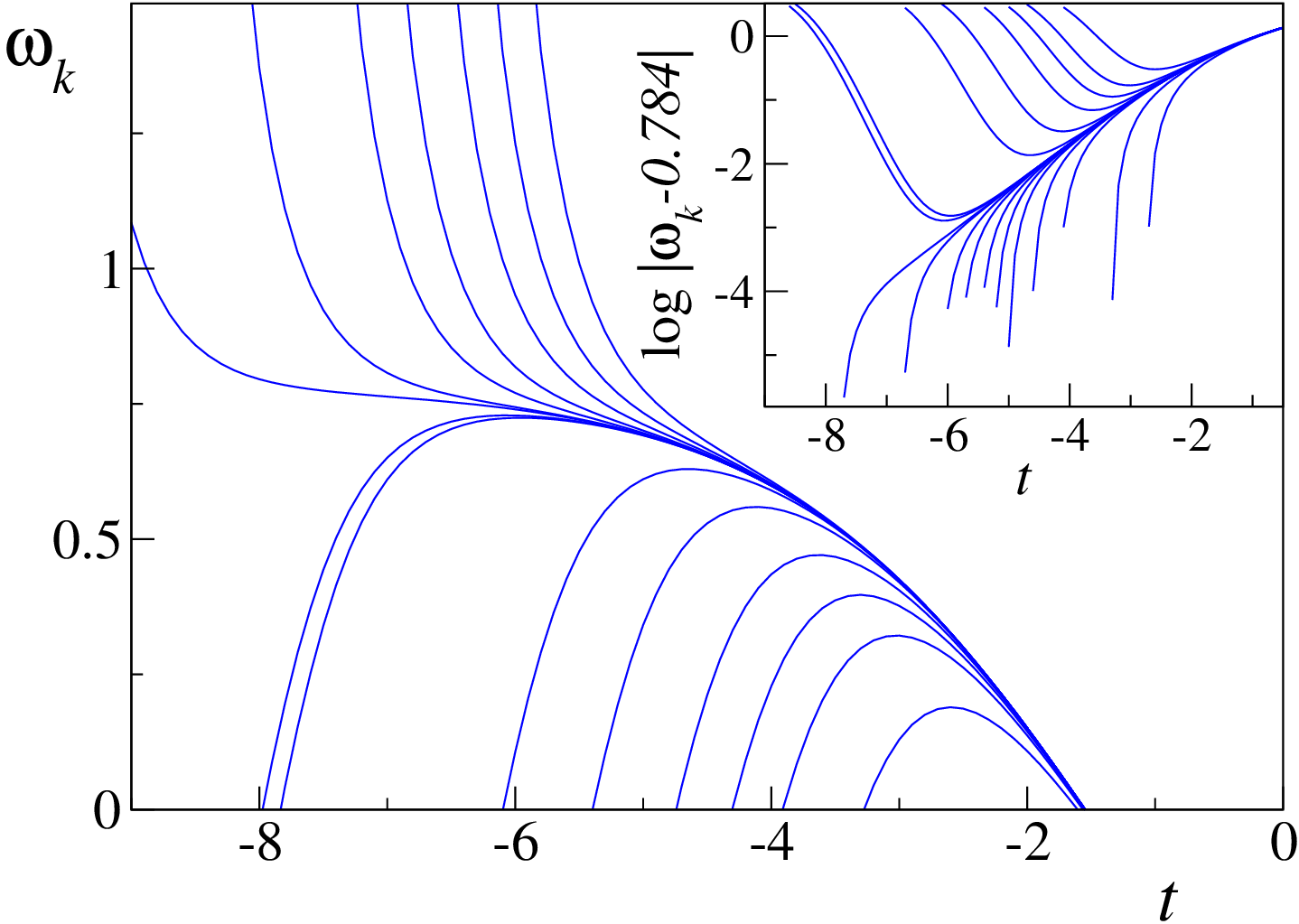}
\end{center}
\caption{\label{nuomega} (color online) Running exponent $\omega_k$ ($d=3$, $N=1$, $\alpha=2.25$, $p_0=0$ and $\rho_0=0$).
Each curve corresponds to a different initial value of $r$ (see Eq.(\ref{cond-init})).
Inset:  the exponential approach to the asymptotic exponent is used to estimate $\omega\simeq 0.784$.}
\end{figure}

The anomalous dimension $\eta$ comes out of the solution of the flow equations, 
which provide a direct estimate of $\eta_k$
 (Fig.~\ref{etafix}). It can also be  extracted from  
$\Gamma_{k=0}^{(2)}(p, \rho=0)\propto p^{2-\eta}$  at small momentum with 
the same result, although this is a much less practical
way.

In the vicinity of the fixed point, the behavior of any dimensionless 
and renormalized quantity, such as 
the dimensionless mass, is as follows (recall that $t=\ln k/\Lambda<0$)
\beq
\label{mass-behavior}
\tilde{m}_k^2=\tilde{m}^2_*+\tilde{m}_1^2 {\rm e}^{-\frac{t}{\nu}}+\tilde{m}_2{\rm e}^{\omega t}+\tilde{m}_3{\rm e}^{\omega_2 t}+\dots
\eeq
with the universal critical exponent $\nu$ describing the departure from the critical surface, and the 
correction to scaling exponents $\omega, \omega_2, \ldots$ describing the initial approach to the fixed point.

In practice, we use the flow of the mass (the flow of $\eta_k$ could also be used) to extract 
$\nu $ and $\omega$ \cite{ref20}. We explore successively regions of $t$-values where one of the exponentials 
in  the equation above dominates. For instance, for $t$ negative enough, we write:
\beq
\log\vert \partial_t \tilde{m}^2\vert \sim -\frac{t}{\nu}+{\rm constant}
\label{slope-nu}
\eeq
to find $\nu$. To extract $\omega$, we choose $\vert t\vert$ large enough but not so large as to leave
the vicinity of the fixed point. We then write
\beq
\log\vert \partial_t \tilde{m}^2\vert\sim \omega t+ {\rm constant}.
\label{slope-omega}
\eeq
Notice that away from the fixed point, the exponents thus determined depend themselves weakly on $t$
since, strictly speaking, (\ref{mass-behavior}) holds only in the infinitesimal vicinity of the fixed point.  
We thus obtain only (slowly) running exponents $\nu_k$ and $\omega_k$.
In practice, these exponents are calculated by taking the $t$-derivative of Eqs.(\ref{slope-nu},\ref{slope-omega}).
The procedure is then repeated for a set of initial conditions that bring the system closer and closer to the critical
point. The estimates of  $\nu_k$ and $\omega_k$ saturate to their fixed point values reflected in the plateau seen 
in Fig.~\ref{nuomega} for  $\omega_k$ ($\omega$ is the most difficult exponent to determine numerically). 
Given such curves, one can further extract even more accurate estimates from the (exponential)
approach to the asymptotic plateau values, see inset of Fig.~\ref{nuomega}.

With this method we could, in principle, extract exponents with almost
arbitrary numerical accuracy. In practice, however, only a few digits are significant: our results suffer indeed
from an uncertainty related to the choice of the cut-off function (see next subsection); besides, it is not necessary
 to present results  with an accuracy that far exceeds the deviation from those with which they are compared.

\subsection{Dependence on renormalization point and regulator}
\label{PMS}

Although as explained above the values of the critical exponents should in principle
 depend neither on the normalization 
point $(\tilde p_0,\tilde\rho_0)$
nor on the shape of the cut-off function $R_k(q)$ this is no longer the case
once approximations are performed. 

In practice, we apply  the  ``principle of minimal sensitivity'',
searching  for  a  local extremum  of  the  physical
quantities under  study \cite{Stevenson,Canet03} in  a ``reasonable'' subspace  of values taken  by $\alpha$
(the parameter  of our cut-off  function (\ref{cut-off-exp})), $\tilde
p_0$,  and  $\tilde\rho_0$. It  is then expected  that the
corresponding  values are  ``optimal'' in  the sense  that  they show,
locally, the weakest dependence on the above parameters.

\begin{figure}[th]
\begin{center}
\includegraphics[width=\columnwidth,clip]{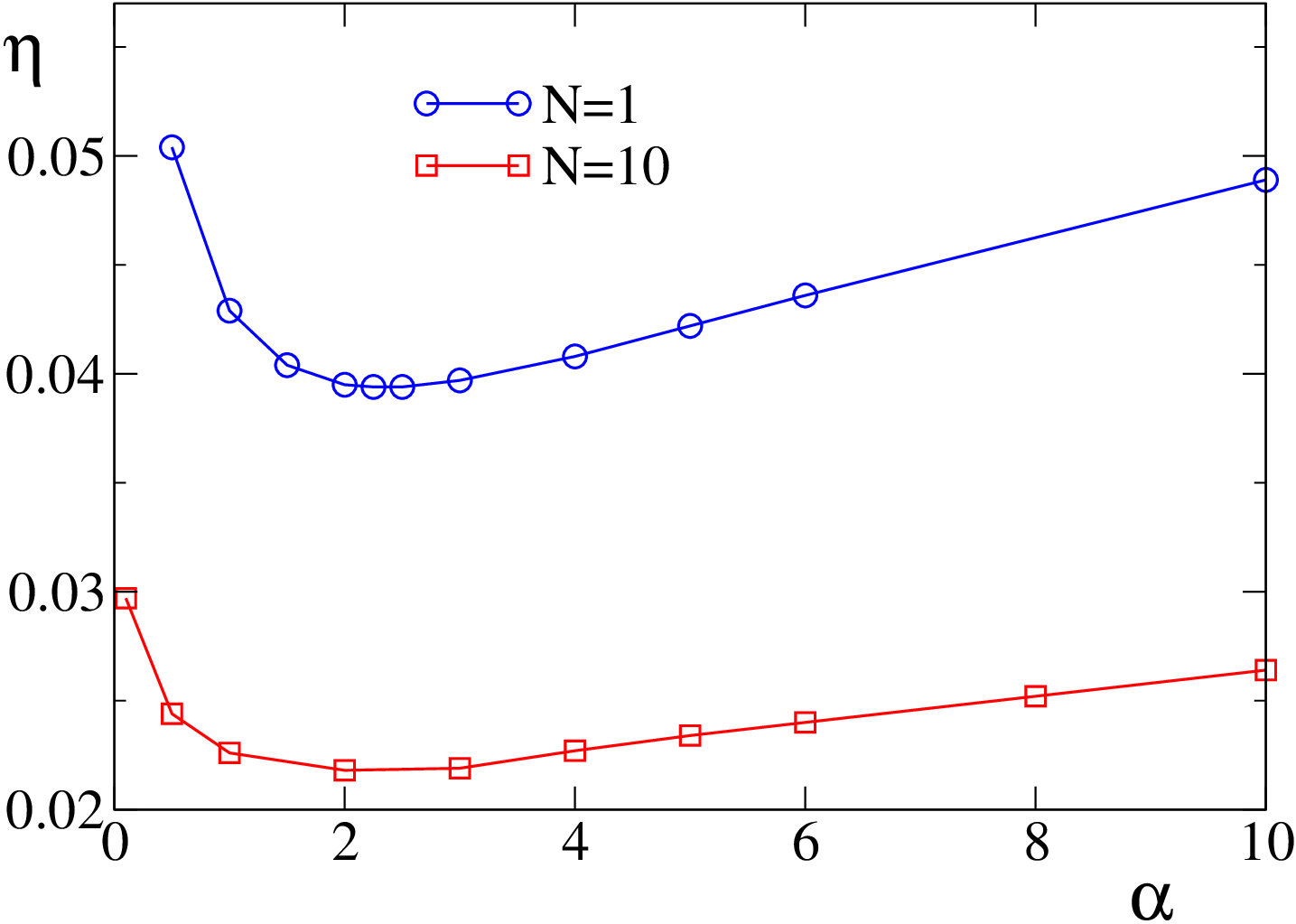}
\end{center}
\caption{\label{etaalpha}(color online) $\eta$ as
a function of the cut-off parameter $\alpha$, for $N=1$ and $N=10$ ($d=3$, $\tilde\rho_0=0, \tilde p_0=0$).}
\end{figure}

Here, we first notice that at fixed $\alpha$ and  $\tilde\rho_0$, the dependence of our estimates on $\tilde p_0$
is  much weaker than that found by varying  $\alpha$ and  $\tilde\rho_0$. 
Fig.\ref{etaalpha} shows the variation of the anomalous dimension  $\eta$ with $\alpha$ for two typical
values of $N$ in three dimensions. 
As in all other cases studied, we observe the existence of a 
unique extremum. 
In the following, we always use these extremum values to report our best estimates for the critical exponents.
Note that we do not show the variations of the exponents with $\tilde\rho_0$ as
they can be shown to be equivalent to those with $\alpha$ \cite{TBP}.
%%%%%%%%%%%%%%%%%%%%%%%%%%%%%%%%%%%%%%%%%%%%%

\subsection{Results for the critical exponents}
\label{critical-exponents}

We now present our results for the critical exponents of the scalar $O(N)$ models in $d=3$. 
They have been obtained with a two-dimensional grid in $\tilde\rho$ and $\tilde q$ with
 $n_\rho=51$ points in the $\tilde\rho$-direction,  $n_q=60$ points in $\tilde q$ and with $\tilde q_{max}=4$, 
$\tilde p_{max}=6$, and a $\tilde\rho_{max}=5 N$. Tables \ref{table_eta}, \ref{table_nu} and \ref{table_w} contain 
our results for the critical exponents  $\eta$, $\nu$ and $\omega$,
together with some of the best estimates available in the literature, 
obtained either from Monte Carlo or resummed perturbative calculations (that we refer to as field theory (FT)). 
Our numbers are all given for the optimal values $\alpha^*$ of the cut-off parameter, 
and the digits quoted remain stable when 
$\alpha$ varies in the range $[\alpha^*-{1}/{2},\alpha^*+{1}/{2}] $.
The quality of these numbers is obvious: 
our results for $\nu$ agree with previous 
estimates to within less than a percent, for all $N$; as for the values of $\eta$ and $\omega$, they 
are typically at the same distance from the 
Monte-Carlo and high temperature series estimates 
(for instance, for $N=1$, $\nu=0.6298(3)$ \cite{arisue2004}) 
as the results from resummed perturbative calculations. 
Our numbers also compare favorably with those obtained at order 
$\nabla^2$ in the DE scheme \cite{Canet03}.

%%%%%%%%%%%%%%%%%%%%%%%%%%%%%%%%%%%%%%%%%%%%%%%%%%%%%%%%%%%%%%%%%%%%%%%%%%

\begin{table*}[ht]
\caption{\label{table_eta}
Results for the anomalous dimension $\eta$ in $d=3$, 
compared with results obtained within the DE at order $O(\nabla^2)$, 
field theory (FT) and Monte Carlo (MC) methods}
\begin{ruledtabular}
\begin{tabular}{lllll}
$N$   &  BMW     &     DE                    &  FT                      & MC  \\
\hline
0     & 0.034    & 0.039\cite{Gersdorff00}   & 0.0272(3)\cite{Suslov08}  & 0.0303(3))\cite{grassberger}  \\
1     &  0.039   & 0.0443\cite{Canet03}      & 0.0318(3) \cite{Suslov08} & 0.03627(10) \cite{hasenbusch10}   \\
2     &  0.041   &0.049\cite{Gersdorff00}    & 0.0334(2) \cite{Suslov08} & 0.0381(2)\cite{Campostrini06} \\
3     &  0.040   & 0.049\cite{Gersdorff00}   & 0.0333(3) \cite{Suslov08} & 0.0375(5)\cite{Campostrini01}  \\
4     &  0.038   & 0.047\cite{Gersdorff00}   & 0.0350(45) \cite{Guida98}   & 0.0365(10)\cite{Hasenbusch01}  \\
10    & 0.022    & 0.028\cite{Gersdorff00}   & 0.024 \cite{Antonenko98}  & - \\
100   & 0.0023   & 0.0030\cite{Gersdorff00}  & 0.0027 \cite{Moshe03}     & - \\
$\mathcal{O}(1/N)$ & $0.23/N$ &                 & $0.270/N$ \cite{Moshe03}    & -
\end{tabular}
\end{ruledtabular}
\end{table*}

\begin{table*}[ht]
\caption{Results for the critical exponent $\nu$ in $d=3$, compared with results obtained within the DE at order $O(\nabla^2)$, 
field theory (FT) and Monte Carlo (MC) methods}
\begin{ruledtabular}
\begin{tabular}{lllll}
$N$   &   BMW         &  DE                        &  FT                       & MC \\
\hline
0     & 0.589         & 0.590\cite{Gersdorff00}    & 0.5886(3) \cite{Suslov08} & 0.5872(5) \cite{Pelisseto07} \\
1     &  0.632        & 0.6307\cite{Canet03}       & 0.6306(5) \cite{Suslov08} &  0.63002(10) \cite{hasenbusch10} \\
2     & 0.674         & 0.666\cite{Gersdorff00}    & 0.6700(6) \cite{Suslov08} & 0.6717(1) \cite{Campostrini06} \\
3     & 0.715         & 0.704\cite{Gersdorff00}    & 0.7060(7) \cite{Suslov08} & 0.7112(5)\cite{Campostrini01} \\
4     & 0.754         & 0.739\cite{Gersdorff00}    & 0.741(6)\cite{Guida98}    & 0.749(2)\cite{Hasenbusch01} \\
10    & 0.889         & 0.881\cite{Gersdorff00}    & 0.859  \cite{Antonenko98} & - \\
100   & 0.990         & 0.990 \cite{Gersdorff00}   & 0.989\cite{Moshe03}       & - \\
$\mathcal{O}(1/N)$ & $1 - 1.034/N$ &                 & $1-1.081/N$ \cite{Moshe03}  & -
\end{tabular}
\end{ruledtabular}
\label{table_nu}
\end{table*}

\begin{table}[ht]
\caption{\label{table_w}
Results for the correction to scaling exponent $\omega$ in $d=3$ compared with results obtained within the BMW method,
field theory (FT) and Monte Carlo (MC) results}
\begin{ruledtabular}
\begin{tabular}{llll}
$N$ & BMW &    FT                    & MC
\\ \hline
0 & 0.83  & 0.794(6) \cite{Suslov08} &  0.88 \cite{grassberger} \\
1 & 0.78  & 0.788(3) \cite{Suslov08} &  0.832(6) \cite{hasenbusch10}  \\
2 & 0.75 & 0.780(10) \cite{Suslov08} & 0.785(20) \cite{Campostrini06}\\
3 & 0.73 & 0.780(20) \cite{Suslov08} & 0.773 \cite{Campostrini01} \\
4 & 0.72 & 0.774(20) \cite{Guida98} &  0.765  \cite{Hasenbusch01} \\
10 & 0.80 & -  &  - \\
100 & 1.00 & - &  -
\end{tabular}
\end{ruledtabular}
\end{table}

In the limit of large $N$, the BMW scheme becomes exact for the 2-point function
for $s\ge 2$ \cite{Blaizot:2005xy,Blaizot:2005wd}. 
This generalizes the fact, shown in \cite{D'Attanasio97}, that the LPA 
($s=0$) is exact in the large $N$ limit for the effective potential. 
It can be verified  from the tables 1 to 4  that  the large $N$ limit 
values $\eta=0$, $\nu=1$ and $\omega=1$  are
approached for large values of $N$. 

We can also perform a $1/N$ expansion \cite{ZINN,Moshe03}. This was already 
done in \cite{BMWnum}, where the BMW scheme was further approximated by the 
use of LPA propagators. As the LPA becomes exact in the large $N$ limit, 
these results are unchanged at first order in $1/N$, except for the use of 
another type of regulator profile. An analytical study of the BMW equations 
in this limit provides the following values for the critical exponents at order $1/N$:  
$\eta={0.23}/{N}$ and $\nu = 1 - {1.034}/{N}$, to be compared with the exact results \cite{Moshe03} 
$\eta={0.27}/{N}$ and $\nu = 1 - 1.081/N$. In \cite{BMWnum} the use of another 
regulator profile allowed us to obtain somewhat better results for $\eta$ 
in this limit: $\eta={0.25}/{N}$. Notice that all 
these analytical results are recovered in our  numerical solution for 
large values of $N$ (notice in fact that terms of order $1/N^2$ are very small already for $N > 4$).

The two-dimensional case, for which exact results exist,  
provides a very stringent test of the BMW scheme. We focus here on the Ising 
model $N=1$ which exhibits a standard critical behavior in $d=2$, and 
the corresponding critical exponents. Notice that the perturbative method that works well in $d=3$ fails here:
for instance, the fixed-dimension expansion that provides the best results in 
$d=3$ yields, in $d=2$ and at five loops, $\eta=0.145(14)$ \cite{pogorelov07} in  
contradiction with the exact value $\eta={1}/{4}$\footnote{It has been conjectured (see \cite{Peli-rev} and
references therein), and this is 
confirmed by $1/N$ calculations, that the presence of non-analytic terms in
the flow of the 
$\phi^4$ coupling $u$ could be responsible for the discrepancy between exact
and perturbative 
results in $d=2$. According to Sokal, no problem should arise when  all 
couplings, including the irrelevant ones, are retained in the RG flow, as done
here. 
This probably explains the quality of our results in $d=2$.}.
We find instead $\eta=0.254$, $\nu=1.00$
in excellent agreement with the exact values $\eta={1}/{4}$, $\nu=1$. 
A more detailed study of $O(N)$ models in $d=2$, at and out of criticality, 
will be presented in a separate work.

\subsection{The function $\Gamma^{(2)}$ at criticality and further tests at intermediate and large momenta} 
\label{section_gamma2}

\begin{figure}[tp]
\begin{center}
\includegraphics[width=\columnwidth, clip] {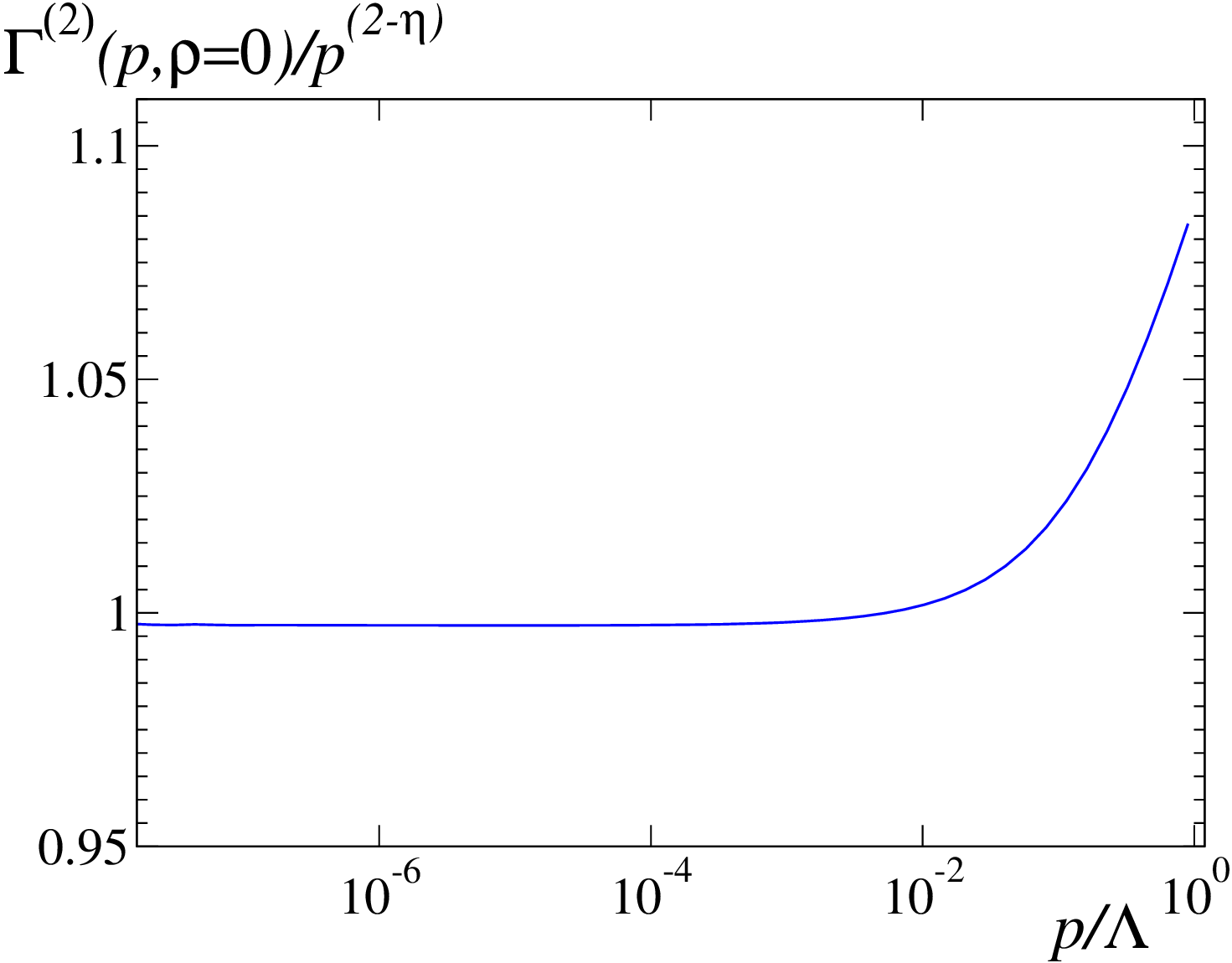}
\end{center}
\caption{\label{selfIR} (color online) The ratio of the 2-point function 
$\Gamma^{(2)}(p,0)$ and of $p^{2-\eta}$ at criticality 
as a function of $p/\Lambda$ ($d=3, N=2, \alpha=2, p_0=0, \rho_0=0$).
The normalization has been chosen so that this ratio starts close to 1 at small $p$. 
The bare dimensionless coupling is $\tilde u=6. 10^{-2}/N$. }
\end{figure}

We now study the momentum dependence of the two-point function at criticality.
In dimension three, the bare coupling constant
$u$ has the dimension of a momentum and thus sets a scale (the Ginzburg length: $\xi_G\sim u^{\frac{1}{d-4}}$). There are typically three momentum domains for
 $\Gamma^{(2)}(p,\rho=0)$ \cite{BMWnum,Benitez09}: 

(i)  the infrared domain defined by $p\ll u$ where $\Gamma^{(2)}(p)\sim u^\eta p^{2-\eta}$. 
We show in Fig. \ref{selfIR}
that this behavior is well reproduced by our solution of the flow equation. To clearly see this regime
on a large range of momentum we have integrated the flow with a bare value of $u$ not too far from the value of 
$\Lambda$: $\tilde u=6. 10^{-2}/N$.

%%%%%%%%%%%%%%%%%%%%%%%%%%%%%%%%%%%%%%%%%%%%%%%
%%%%%%%%%%%%%%%%%%%%%%%%%%%%%%%%%%%%%%%%%%%%%%%%

\begin{figure}[tp]
\begin{center}
\includegraphics[width=\columnwidth, clip] {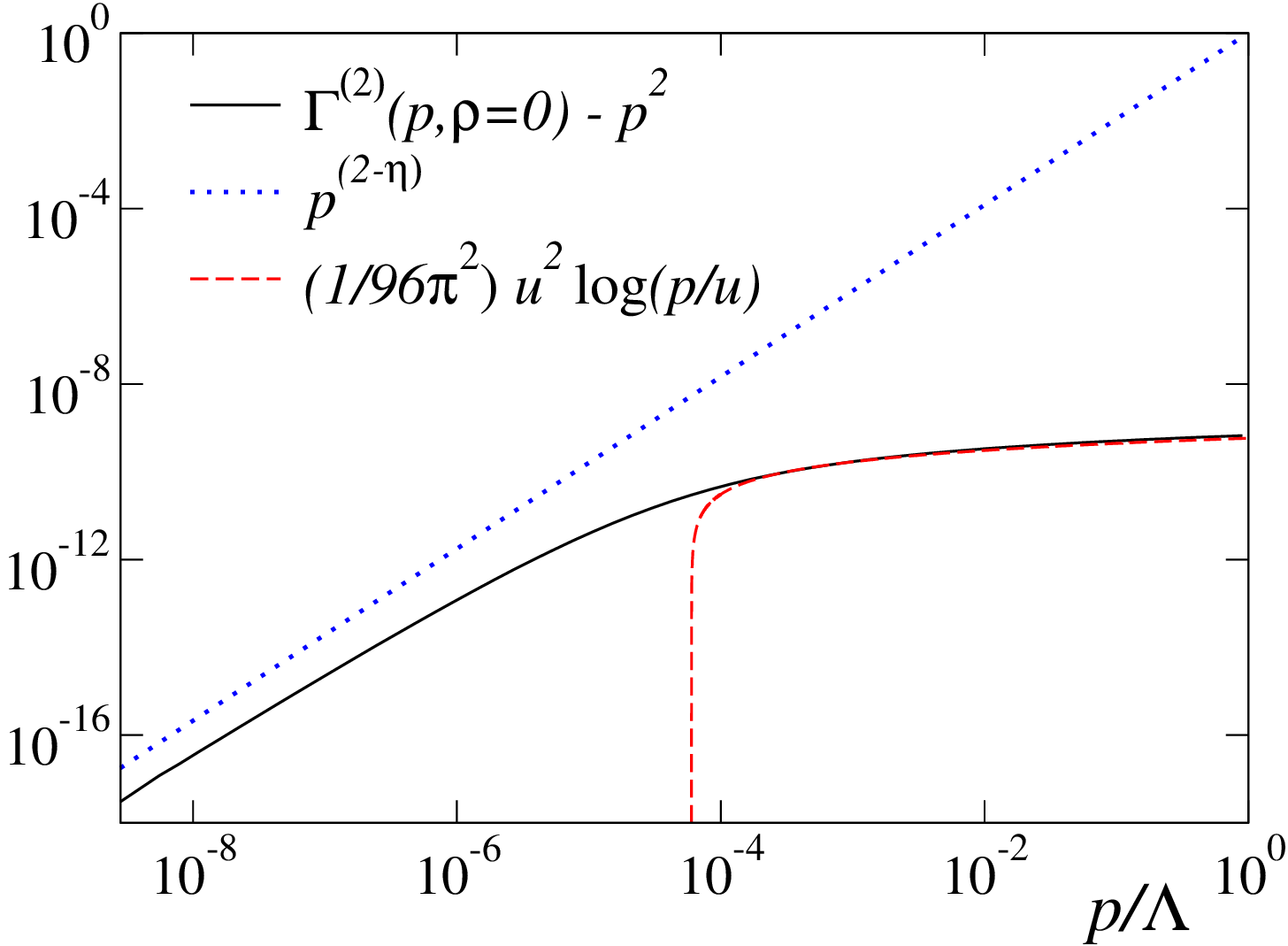}
\end{center}
\caption{\label{selfUV} (color online) The  difference $\Gamma^{(2)}(p,0)-p^2$ at criticality as a 
function of $p/\Lambda$, compared with its 
expected ultra-violet behavior $\sim u^2 \log p/u$ ($d=3, N=2, \alpha=2, p_0=0, \rho_0=0$).
The  infrared $p^{2-\eta}$ behavior is also shown (see text). The bare 
dimensionless coupling is  $\tilde u= 10^{-6}/N$. }
\end{figure}

(ii) the ultra-violet domain defined by $p\gg u$ (and $\Lambda\gg p$) where 
$\Gamma^{(2)}(p)$ can be studied perturbatively
and is found to behave at two loops as  $\Gamma^{(2)}(p)-p^{2}\sim  - 
(\mathcal{C}_N /{96\pi^2})u^2\log p/u$ (with $\mathcal{C}_N=(N+2)/3$). 
It was shown in \cite{BMWnum} that in the BMW approximation, and at 
large momenta $\Delta(p,\rho=0)$ behaves as $u^2\ln(p/u)$,  more precisely,
\begin{equation}
\label{finLO}
\frac{\partial \Delta_{k=0}(p,0)}{\partial|p|} = \mathcal{C}_N \frac{u^2}{2|p|}
\int_{l,q} \partial_t R_k(l)G_0^2(l)  G^2_0(q).
\end{equation}
The $u^2\log p/u$ behavior is thus retrieved, see Fig.\ref{selfUV}, with a prefactor that however depends on $R_k(q)$.
With the exponential cut-off function, Eq.(\ref{cut-off-exp}), the prefactor can only be calculated numerically. 
We have studied its dependence on $\alpha$ and shown that there is an extremum around $\alpha \sim 5$
where the difference with the exact result is about 8$\%$. Of course, 
this UV behavior shows up only if the bare coupling $u$ is sufficiently small compared to $\Lambda$. We have 
chosen $\tilde u=10^{-6}/N$ to have a large UV domain where this behavior is clearly seen. 
Note that at small $p$, $\Gamma^{(2)}(p)- p^{2}\sim  p^{2-\eta}$ which is visible on  
Fig.\ref{selfUV} although this regime is 
approached very slowly.

%%%%%%%%%%%%%%%%%%%%%%%%%%%%%%%%%%%%%%%%%%%%%%%%
%%%%%%%%%%%%%%%%%%%%%%%%%%%%%%%%%%%%%%%%%%%%%%%%%%%%

(iii) the cross-over between the infrared and ultra-violet domains. This regime of momentum is visible
on both figures \ref{selfIR} and \ref{selfUV} for $p\simeq u$.

\begin{table}
\caption{\label{table_c}
Results for the quantity $c$ defined in the text.}
\begin{ruledtabular}
\begin{tabular}{llll}
 $N$ & BMW & lattice & $7$ loops \cite{Kastening03}  
\\ \hline
1 &  1.15 & 1.09(9)\cite{Sun02}  & 1.07(10)   \\
2  & 1.37 & 1.32(2) \cite{Arnold01} &  1.27(10) \\
& & 1.29(5)\cite{Kashurnikov01} &       \\
3 & 1.50 &  & 1.43(11)   \\
4 &  1.63  & 1.60(10)\cite{Sun02} & 1.54(11)   \\
10 &  2.02 &  &    \\
100 &  2.36 &  &    
\end{tabular}
\end{ruledtabular}
\end{table}

For purposes of probing the intermediate momentum region between the IR and the UV, we have
calculated the quantity
\begin{equation}\label{c}
c =   -\frac{256}{uN} \zeta[{3}/{2}]^{-\frac{4}{3}} \,
\!\int\! {\rm d}^3 p \left ( \frac{1}{
\Gamma^{(2)}(p)} \!-\! \frac{1}{p^2}\right)
\end{equation}
which is very sensitive to the cross-over regime: the integrand in
Eq. (\ref{c}) is peaked at $p \sim (N u)/10$
\cite{Blaizot04}. For this reason, the calculation of $c$  has been used as a benchmark for
non-perturbative approximations in the $O(N)$ model.

 In the $O(2)$ case and for $d=3$, this quantity
determines the shift of the critical temperature of the weakly
repulsive Bose gas \cite{Baym99} (notice that $c$ is not defined for $d=2$). It has thus been much studied
recently using various methods, even for other values of $N$. In
particular, the large $N$ limit for this quantity has been
calculated analytically and found to be $c=2.3$
\cite{Baym00}.
In this work, we have found the values for
$c$ for some representative values
of $N$. Our results, compared to the best ones available in the literature (with their
corresponding errors when available), are presented in Table
\ref{table_c}. For all values of $N$ where lattice and/or 7-loops
resummed calculations exist, our results are within the error
bars of those calculations (and comparable to those obtained from 
an approximation specifically designed for this quantity \cite{Blaizot:2005wd,ref41}), except for $N=2$, where very precise
lattice results are available. In the large $N$ limit, one can see
that our results differs from the exact value by less than 3\%.
Notice that the large $N$ behavior of
the quantity $c$ is in fact of order $1/N$ \cite{Baym00}, which as we have seen is not calculated exactly
at this level of the BMW approximation.

Altogether, we can see that the BMW method is able to reproduce the correct behavior of the
2-point function at criticality in all momentum regimes. Note in particular that 
this is not the case of conformal field theoretical methods
that are only able to capture at criticality  the  conformally invariant $ p^{2-\eta}$ behavior
but that can reproduce neither the ultra-violet behavior, corresponding to $u\ll p \ll \Lambda$,
nor the cross-over between the infrared and ultra-violet regions, corresponding to $ p \simeq u$.

\section{Scaling functions}
\label{scaling}

As an approximation of the NPRG, the BMW  scheme allows us to investigate 
all momentum, temperature and external magnetic 
field regimes, and is not restricted to the long distance physics at criticality. 
A particularly interesting, and a priori 
difficult regime is the critical domain, where the correlation length is large but finite. 
In this case, an appropriately 
rescaled two-point function shows a universal behavior. 
As the BMW approximation allows for the calculation of genuine momentum-dependent quantities, the 
calculation of this scaling function and its comparison with the best available 
theoretical results from the literature 
and with experimental data represent one of the most stringent tests of the approximation. 

In this work, we  consider the case $N=1$ relevant, for instance, to describe the critical behavior 
of fluids near the liquid-gas critical point. Near this point and for 
$p\ll \xi_G^{-1}\sim u$ one expects the general scaling behavior
\begin{equation}
 G^{(2)}_\pm(p)=\chi g_\pm(p\xi)
\end{equation}
with, by definition, $G^{(2)}$ the density-density correlation function,  $\chi^{-1}=\Gamma^{(2)}(p=0)$
 the compressibility and $\xi^{-2}=k^2 \tilde{m}^2_k,$ with $k \to 0$,  the 
correlation length that diverges close to criticality with the $\nu$ critical exponent. 
Here $\pm$ refers to the two phases, above and below the 
critical temperature respectively. The functions $g_\pm(x)$, normalized so that
\begin{equation}
 g^{-1}(x)=1+x^2+O(x^4),
\end{equation}
are universal. Their limiting behavior is well known. For small $x$ they are well described by the 
 Ornstein-Zernicke (mean-field) approximation:
\begin{equation}
 g_{OZ}(x)=\frac{1}{1+x^2}.
\end{equation}
The corrections to the Ornstein-Zernicke behavior are usually parameterized as \cite{martin-mayor02}
\begin{equation}
\label{correction-OZ}
 g_\pm(x)^{-1} =1+x^2 + \sum_{n=2} c_n^{\pm} x^{2n} \;.
\end{equation}
The above behavior of $g_\pm(x)^{-1}$ is a priori valid only for $x<1$ but 
since the coefficients $c_n$ are very small, it turns out that 
the Ornstein-Zernicke approximation is actually
valid over a wide range of $x$ values, as we shall see later.
For  large $x$ (that is, $\xi \gg p^{-1}$) the scaling functions show critical 
behavior with an anomalous power law decay
\begin{equation}\label{g-critique}
 g_\pm(x)=\frac{C_1^\pm}{x^{2-\eta}},
\end{equation}
which allows for the experimental determination of the exponent $\eta$. This expression also allows 
for corrections, as given by Fischer and Langer \cite{fisher68}
\begin{equation}
 g_\pm(x)=\frac{C_1^\pm}{x^{2-\eta}} \bigg(1+\frac{C_2^\pm}{x^{(1-\alpha)/\nu}} +\frac{C_3^\pm}{x^{1/\nu}}+\dots \bigg).
\end{equation}

Different approximate results for the universal scaling functions exist in the literature, obtained
either by Monte Carlo methods \cite{martin-mayor02}, or by the use of an analytical ansatz, 
interpolating between the two know limiting regimes (\ref{correction-OZ}) and (\ref{g-critique}), using $\varepsilon$ expansion results 
(the Bray approximation \cite{bray76}). Experimental results from neutron 
scattering in ${\rm CO}_2$ near the critical point also exist \cite{damay98}.

In Bray's interpolation for the high temperature phase one assumes $g^{-1}_+(x)$ to be well 
defined in the complex $x^2$ plane, with a branch cut in the negative real $x^2$ axis, 
starting at $x^2=-r^2_+$, where $r_+^2=9M_{\rm gap}^2\xi^2\equiv 9 S_M$, following the theoretical 
expectation that the singularity of $g_+(x)$ nearest to the origin is the three-particle 
cut \cite{ferrell75,bray76}. The parameter $M_{\rm gap}$ is the mass gap of the 
Minkowskian version of the model. For the $\phi^4$ theory, it is known that
 the difference between the mass gap and $\xi^{-1}$ 
is very small and replacing one by the other corresponds to an error
which is beyond the accuracy of our calculation \cite{martin-mayor02}. Then Bray's ansatz in the 
high temperature phase (the only phase studied in the following) reads:
\begin{eqnarray}
 g^{-1}_+(x)=&&\hspace{-3mm}\frac{2\sin \pi \eta/2}{\pi C_1^+}\nonumber\\
 &&\hspace{-18mm}\times\int_{r+}^{\infty} du \; F_+(u) \bigg[\frac{S_M}{u^2-S_M}+\frac{x^2}{u^2+x^2} \bigg]
\end{eqnarray}
where $F_+(u)$ is the spectral function, which satisfies $F_+(+\infty)=1$, $F_+(u)=0$ 
for $u<r_+$, and $F_+(u)\geq 0$ for $u\geq r_+$. On top of this, one must impose $g^{-1}(0)=1$, 
which fixes the value for $C_1^+$.

One must then specify $F_+(u)$. Bray \cite{bray76} proposed the use of a spectral 
function with the exact Fischer-Langer asymptotic behavior, of the type
\begin{equation}
 F_{+,B}(u)=\frac{P_+(u)-Q_+(u)\cot \frac{1}{2}\pi \eta}{P_+(u)^ 2+Q_+(u)^ 2}
\end{equation}
where
\begin{eqnarray}
 P_+(u)&&\hspace{-3mm} =1+\frac{C_2^+}{u^\iota} \cos\frac{\pi \zeta}{2}+ \frac{C_3^ +}{u^{1/\nu}}\cos \frac{\pi}{2\nu} \nonumber \\
Q_+(u)&&\hspace{-3mm} =\frac{C_2^+}{u^\iota} \sin \frac{\pi \zeta}{2}+ \frac{C_3^ +}{u^{1/\nu}}\sin \frac{\pi}{2\nu} 
\end{eqnarray}
with $\zeta \equiv (1-\alpha)/\nu$. This definition contains a certain number of
 parameters. On top of the critical exponents, which can be injected using either the 
BMW values or the best available results in the literature, one must also fix $S_M^+$, $C_2^+$ 
and $C_3^+$. For $S_M^+$ one can use the best estimate in the literature, given by the 
high temperature expansion of improved models
\cite{Campostrini02b}. Bray proposed to fix $C_2^++C_3^+$ to its $\varepsilon$-expansion 
value $C_2^++C_3^+=-0.9$, and to then determine $C_1^+$ by requiring $F_{+,B}(r_+)=0$. These conditons 
allows for a little parameter tuning, by adjusting the relative weight of the $C_2^+$ and $C_3^+$ parameters.
When comparing our results with Bray's ansatz, we shall use this freedom. We now turn to the scaling function
computed by the BMW method.

\begin{figure}[tp]
\begin{center}
\includegraphics[width=\columnwidth,clip] {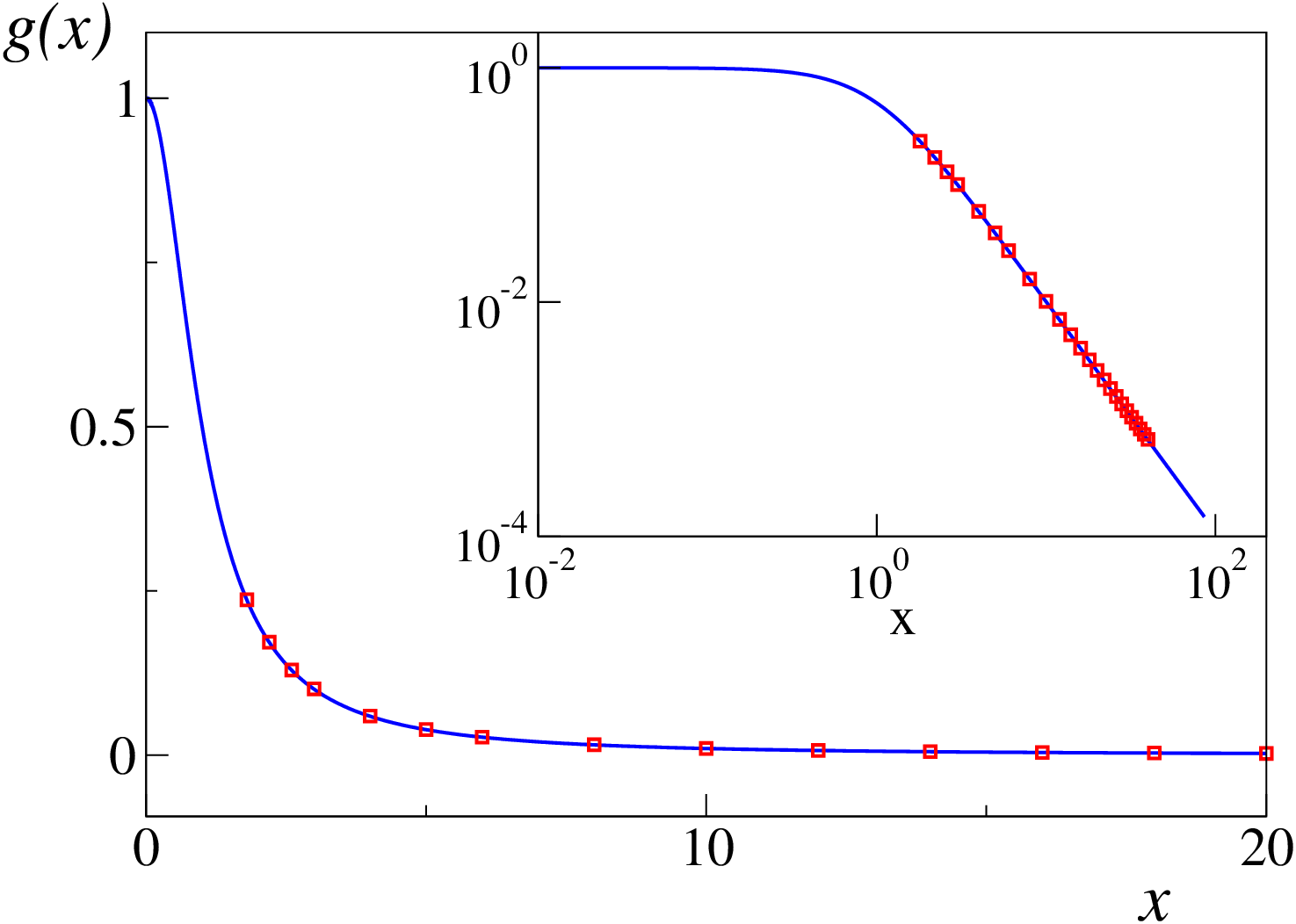}
\end{center}
\caption{\label{scaling-function} (color online) The  2-point scaling function $g(p\xi)$ as a function of $x=p\xi$ 
in the high-temperature phase ($d=3$, $N=1$).
Solid blue line: BMW result. Red squares: Experimental results of \cite{damay98}.
Inset: same data with logarithmic scales.}
\end{figure}

In terms of the variables used in this paper, we find that
\begin{equation}
 g^{-1}(p\xi)=\frac{(p\xi)^2 + \Delta(p\xi,0)+Z_k k^2 m^2_k(0)}{Z_k k^2 m^2_k(0)}
\end{equation}
when $k\to 0$. In this work, 
 for purposes of comparison with existing results, we have only computed the high temperature scaling function.
We have performed the calculation for different values of the correlation length 
(and hence of the reduced temperature). When plotted, one can 
indeed see perfect data collapse for different values of $\xi$, which is the first non trivial test of the quality 
of our results for the scaling function.

\begin{figure}[ht]
\begin{center}
\includegraphics[width=\columnwidth,clip] {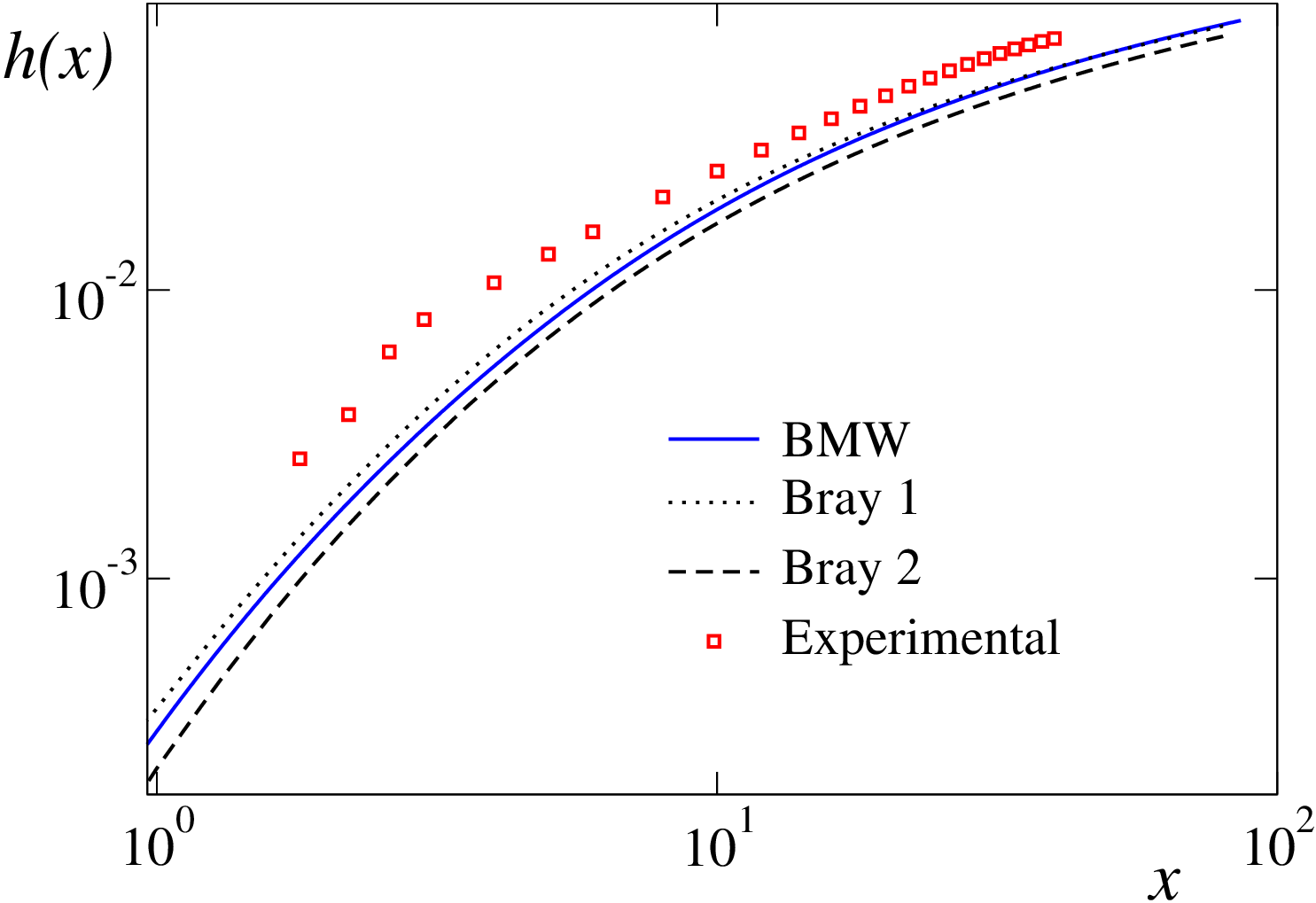}
\end{center}
\caption{\label{h-comparison}(color online) Deviation of the scaling function to its trivial  
Ornstein-Zernicke form, Eq.(\ref{fonction_h}).
The dotted and dashed lines correspond to  two ``extreme'' choices of the  parameters $C_2$ and 
$C_3$ of  Bray's ansatz. Dotted line: $C_1=0.924$, $C_2=1.8$, $C_3=-2.28$.
Dashed line:  $C_1=0.918$, $C_2=2.55$, $C_3=-3.45$.}

\end{figure}

In Fig.\ref{scaling-function} we plot the BMW scaling function together with the experimental 
results from reference \cite{damay98}. 
Due to the small values taken by the coefficients $c_n$ and the critical exponent $\eta$ in $d=3$, 
 the Ornstein-Zernicke behavior dominates even beyond $p\xi=1$. 
In order to measure the deviation from this behavior, one usually makes 
use of the auxiliary function
\begin{equation}
 h(x)=\log \bigg[ \frac{g(x)}{g_{OZ}(x)} \bigg].
\label{fonction_h}
\end{equation}
In Fig.\ref{h-comparison} we plot this function together with the experimental results from \cite{damay98} 
and the results 
from the Bray ansatz for two ``extreme''  choices of the $C_2^+$ and $C_3^+$ parameters. One can there see 
that the BMW approximated 
result compares very well with all these results. In particular, it is in between the results obtained from the 
two Bray ansatz considered. 

Let us  mention that even with large system sizes,
the Monte Carlo results suffer from significant systematic errors for $p\xi$ larger than typically 5 to 10.
This probably comes from the fact that the universal behavior of the structure factor shows up only when
$\xi$ and the separation $l$ between the spins at which we calculate the correlation function are large 
compared to the lattice spacing and small compared to 
the lattice size: even for lattice sizes of a 
few hundreds of lattice spacings this leaves only a small window of useful values of $\xi/l$ \cite{martin-mayor02}.

On top of these results we can also compare results for the values of the 
coefficients $c_2^+$ and $C_1^+$. The results for 
BMW are $c_2^+\sim -4.5 \times 10^{-4}$ to be compared with the IHT best estimate 
\cite{Campostrini02b} $c_2^+= -3.90(6) \times 10^{-4}$, 
whereas for $C_1^+$ BMW yields $C_1^+=0.914$, to be compared with the $\varepsilon$-expansion result $C_1^+=0.92$.

We conclude this section by noting that (i)  the structure factor encompasses much more
informations on the universal behavior of a model than the (leading) critical exponents (that are moreover difficult
to measure experimentally), (ii)  Bray's ansatz, although powerful, depends on
two parameters $C_2$ and $C_3$ that are poorly determined perturbatively as well as on two critical exponents,
(iii)  the present state of the art of the Monte Carlo simulations is by far insufficient to compute 
reliably the structure factor in the interesting region of momentum where $p\xi$ is large, (iv) the BMW method
leads to a determination of the structure factor that has no free parameter once a choice of regulator
has been made (possibly involving an optimization procedure as described in Section \ref{PMS}). The results above,
summarized in Fig.\ref{h-comparison}, suggest that the BMW method leads to an accurate  determination
of the structure factor in the whole momentum range while the experimental results seem to suffer at small
momentum from  systematic deviations. 

\section{Relation with the derivative expansion}
\label{sectionDE}
The validity of the DE  is rarely questioned, satisfactory results being taken as 
an a posteriori check.
We show now that the BMW approach allows for  a 
deeper understanding of its range of applicability, and of some of its peculiar features.

The ansatz defining the order of the DE (see, for instance, Eq.(\ref{ordre2})  for the order 2)
is used to

(i) define the quantities to be determined, which, in the case of order 2, are the effective potential and the field normalization, both functions of the (constant) field $\phi$
\begin{equation}
\left\{
\begin{array}{l}
\displaystyle{V_k(\phi)= \frac{1}{\Omega} {\Gamma_k[\phi(x)]}{\big\vert_{\phi(x)=\phi}} }\\ \\

\displaystyle{Z_k(\phi)=\partial_{p^2}\left(\Gamma_k^{(2)}[p,\phi] \right){\big\vert_{p=0,\phi}}}\,.
\end{array}
\right. 
\label{UZ}
\end{equation}

(ii) compute the  $n$-point functions $\Gamma_k^{(n)}$ and  the propagator $G_k=(\Gamma_k^{(2)}+R_k)^{-1}$  
that enter the  right hand sides of the  flow equations of $V_k$, $ Z_k$, etc. 

In short, the DE projects the functional $\Gamma_k[\phi]$ on a polynomial expansion in powers 
of the derivatives of the field, the expansion coefficients being field dependent. 
In Fourier space, the DE amounts to a polynomial expansion of the $n$-point functions
$\Gamma_k^{(n)}(p_1,\cdots,p_n,\phi)$ 
in powers of the momenta $p_i$, around vanishing momenta (see for instance Eq.~(\ref{Vgenerating})). 
At this point, it is useful to introduce a distinction between {\it external} momenta, the momenta that
 appear in the $n$-point function $\Gamma_k^{(n)}(p_1,\cdots,p_n,\phi)$ whose flow is being 
considered, and the {\it internal} momentum, denoted by $q$,
 appearing in the $n$-point functions in the  r.h.s. of the corresponding flow equation and which is
integrated over. In contrast to what is done 
in the BMW approximation, in the DE no distinction is made between these two sets of momenta, which 
can lead to  inconsistencies. For  instance, in  the flow equation  for $Z_k(\phi)$ at order 2,
the product $\Gamma_k^{(3)}(p,q,-p-q) \Gamma_k^{(3)}(-p,-q,p+q)$ (see Eq.(\ref{gamma2champnonnul}))
leads to four terms of order four: $(p^2)^2,p^2 q^2, (p.q)^2,(q^2)^2$, that, in a strict expansion to this order, 
should be neglected (note that this is not what is  usually done in the DE context).
In fact, since $Z_k(\phi)$ is already the coefficient of the  $p^2$ term in the expansion 
of $\Gamma_k^{(2)}(p,\phi)$, any dependence of $\Gamma_k^{(3)}$ (and of  $\Gamma_k^{(4)}$) 
on the internal momentum $q$ should  be neglected in $\partial_kZ_k(\phi)$ at this order of the DE. 
Since the BMW approximation at order $s=2$ precisely consists in setting $q=0$ in $\Gamma_k^{(3)}$ and 
 $\Gamma_k^{(4)}$ in the flow equation of $\Gamma_k^{(2)}$, we conclude that at this order
 the BMW approximation  contains all terms of the DE at order $\nabla^2$.

The BMW approximation, that disentangles 
the roles of the internal and external momenta, differs deeply from the DE precisely on the point explained above: 
as the DE, it takes advantage
of the fact that the internal momentum is cut-off by $\partial_kR_k(q)$ in order to expand in powers 
of $q/k$ (in fact only the leading term, $q=0$, is retained), but does not rely on the smallness
of the external momenta.

In fact, the natural expansion parameter of the DE
is the ratio $p/k$ or $p/m$, whichever is smallest, where $m$ is the smallest of the masses 
that may appear in the problem considered:
When $k$ is much larger that all  masses, these can be ignored and $p/k$ is the expansion parameter.
When $k$ becomes smaller that the smallest mass,  the flow
essentially stops and the expansion parameter becomes  $p/m$ in the limit $k \to 0$.
Thus, it is plausible that
the DE performed as a power series in $p/k$ in a critical theory ($m=0$) possesses  
a radius of convergence of the same order as the DE performed 
as a power series  in $p/m$ in a massive theory at $k=0$.  
In this last case, the radius of convergence is known for $N=1$ in dimension three (\cite{ferrell75,bray76}):
It is 3 in the symmetric 
phase and 2 in the broken phase
\footnote{The reason is easily understood in the Minkowskian version of the theory. In this case,
$3m$ is the particle production threshold (in the symmetric phase) that reflects itself as 
a pole in the complex momentum plane.
This pole, which is the closest to the origin, determines the radius of convergence of the expansion
in powers of $p$. The same reasoning leads to $2m$  in the broken
phase except if there exists a two-particle bound state, in which case its mass (which is smaller than
$2m$) determines the radius of  convergence. It is very probable  
that  such a bound state exists in the $d=3$, $N=1$ case \cite{Caselle02}
and its mass has been found of order $1.8m$. Note that this kind of analysis can be generalized to any model 
with a Minkowskian unitary extension.}.

The above arguments suggest that the DE is not able to describe $k$-dependent correlation 
functions with external momenta higher than typically $3\,\mathrm{max}(k,m)$. 
In particular, in the critical case where massless modes are present, the DE is only suited for 
the calculation of physical (that is at $k=0$) correlation functions at $p=0$: The anomalous momentum behaviour 
$\Gamma_{k=0}^{(2)}(p)\sim p^{2-\eta}$, 
valid at small $p$, will not emerge  
at any order of the DE. Of course, this does not mean that the anomalous 
dimension cannot be determined within the DE, as one can exploit general scaling relations 
and the fact that the anomalous dimension enters also quantities that are defined {\it at} zero 
momentum. Thus, for instance, $\eta$ can be estimated from the $k$-dependence of 
the normalization factor  $Z_k\sim k^{-\eta}$
(or alternatively from the large field behavior of the fixed-point dimensionless effective 
potential). In contrast, the BMW approximation correctly captures the anomalous scaling 
of $\Gamma_{k=0}^{(2)}(p)$ at small $p$, 
and this is a direct consequence of the fact that  no expansion 
in external momenta is performed\footnote{We recall 
that the two determinations of $\eta$ performed within the BMW 
approximation either through the momentum dependence
of $\Gamma_{k=0}^{(2)}(p)$ or from $Z_k$ lead to the same values of this exponent. }. 

The origin of the difficulties of the DE is that it
does not have good decoupling properties in the momentum range $p\gg k$. The decoupling property, crucial
for universality, means, on the example on the 2-point function, that $\Gamma_{k}^{(2)}(p)$ 
becomes almost $k$-independent when $k\ll p$ and that
therefore $\Gamma_{k=0}^{(2)}(p)\simeq \Gamma_{k=p}^{(2)}(p)$. One 
could thus naively expect that  external momenta $\{ p_i\}$, $i=1,\dots, n$ 
play the role of  infrared regulators in the flow of $\Gamma_{k}^{(n)}(\{ p_i\})$ and that
 when $k< p_i, \forall i$ the flow of $\Gamma_{k}^{(n)}$ (almost) stops.
In fact,  in flow equations, external momenta play, at best, 
the role of infrared regulators when all momenta involved
(external and internal) are not in an exceptional configuration. The problem is that even
when the external momenta are not exceptional, the integral over the internal momentum $q$ in the flow
equation of  $\Gamma_{k}^{(n)}$  involves
vertex functions ($\Gamma_{k}^{(n+1)}$ or $\Gamma_{k}^{(n+2)}$) in exceptional  configurations. Depending
on the approximation scheme, this can spoil the decoupling property that, undoubtly,  should hold  
for the (physical, that is $k\ll p$) correlation 
functions themselves when they are evaluated in non-exceptional configurations. 
The difficulty is therefore
to devise an approximation scheme that satisfies the decoupling property. While this is the case
of the BMW scheme it is  neither of perturbation theory nor of the DE.  
One can nevertheless try to extract from the DE the gross behavior of $\Gamma^{(2)}(p)$ 
(and of the other functions) by
stopping by hand the flow at $k=p$ and identifying $\Gamma_{k=p}^{(2)}(p)$ with
 $\Gamma_{k=0}^{(2)}(p)$. This idea has been explored in \cite{Blaizot:2005wd} 
(see also \cite{Dupuis2009}). The resulting 
correlation functions roughly show the expected momentum behavior, but as analyzed in detail in 
\cite{Blaizot:2005wd}, it 
does not seem  possible to extend this first qualitative analysis and to  obtain  quantitatively 
precise correlation functions without having recourse to BMW.

To gain further insight into the validity of the DE, we may consider a simple analytical 
representation of the 
function $\Gamma_{k=0}^{(2)}(p,\phi)$ determined with the BMW approximation at order
$s=2$, which, as we have shown,  is very close to the exact 2-point function over the whole momentum range.
The following formula (inspired by Eq.(2.33) of \cite{Berges02})
\begin{equation}
\begin{array}{ll}
\Gamma_{k}^{(2)}(p,\rho)= &A p^2\left( p^2+ b\, k^2+ b' \,M^2_k(\rho)\right)^{-\eta_k/2} \\
&+ V_k'+2 \rho V_k''
\end{array}
\end{equation}
where $A$, $b$, and $b'$ are independent of $p$ and $\rho$, and  $M^2_k(\rho)$ is a function
homogeneous to a square mass,  provides a good fit of the BMW results when $k$, as well
as $M^2_k(\rho)$,  are very small compared to the ultra-violet 
cut-off $\Lambda$. 
This formula encompasses the two different regimes that characterize the behavior of $\Gamma_{k}^{(2)}(p)$ at small $p$:
First, for  $p$ small compared to $\Lambda$ and
large compared to $k$ and to the mass, it yields $\Gamma_{k}^{(2)}(p)\sim p^{2-\eta_k}$,
with $\eta_k$ the running anomalous dimension. Thus,  the critical behavior is captured 
for $k$ sufficiently small for $\eta_k$ to be quasi-stationary and (almost) equal to $\eta$.
Second,
for $p$ small compared  to either $M_k$  or $k$, one can expand 
$\Gamma_{k}^{(2)}(p,\phi)$ in powers of $p^2/(k^2+M_k^2)$ and get:
\begin{multline}
\Gamma_{k}^{(2)}(p,\rho)= A' \left(k^2+ b'' \,M^2_k(\rho)\right)^{-\eta_k/2} \\
\times p^2\left( 1+f_{1,k}(\rho) p^2+f_{2,k}(\rho) p^4+\dots\right)\\
+ V_k'+2 \rho V_k''.\ \ \ \ \ \ \ \ 
\label{ansatz_DE}
\end{multline}
This is the kind of ansatz considered by the DE and it illustrates how the anomalous dimension can 
be extracted from the $k$-dependence of the coefficient of the $p^2$ term in the running 
action \cite{Blaizot:2005wd}.

Finally, let us stress that the above remarks, while they provide some 
justification for the DE and in particular specify the  conditions for 
its validity, are not  sufficient to prove convergence, which may be 
strongly affected by the regulator.  In particular, one may expect 
systematic errors in cases where the range of the cut-off function 
$R_k(q)$  is not  smaller than the natural radius of convergence of the DE. 
Notice however that at least for  $N=1$  in $d=3$, the smallness of the $c_n$ 
coefficients in Eq. (\ref{correction-OZ}) suggest that even at low order
the DE should be able to capture  the low momentum physics.
An in-depth study of this issue will be presented in \cite{TBP}.

\section{Conclusions}
\label{Conclusions}
In this paper we have presented the complete numerical implementation of the BMW approximation scheme
that allows for a solution of the NPRG flow equations  keeping the full momentum dependence of
the 2-point function. At the level considered in this paper, this amounts to solve two coupled equations
for the effective potential and the 2-point function. These equations can be solved by elementary numerical
techniques. 

We have considered applications to the $O(N)$ models, mostly in dimension $d=3$.
An  accurate momentum dependence of the 2-point function has been obtained from the low momentum
critical region to the high momentum, perturbative, region (such a region exists when the dimensionful bare
coupling is small compared to the ultraviolet cutoff). In particular, the critical exponents are accurately
determined as was already reported in \cite{Benitez09}. The additional results presented in this paper
concerns the scaling functions which probe a different aspect of the momentum dependence of the
2-point function in the vicinity of the critical point. We have considered more specifically the scaling
function for the case $N=1$ above the critical point and have shown that it is in excellent agreement
with the best available theoretical estimates. Interestingly, these estimates, including ours, differ significantly
from the experimental data at small momenta. These scaling functions, which are difficult to obtain with
other more conventional techniques, including Monte Carlo simulations, come out directly from the 2-point
function obtained by solving the flow equations. 

Another information of physical interest which is also contained in the
2-point function that we compute is its field dependence. Thus a natural application of the present
method could be the investigation of the $O(N)$ models in the presence of an external magnetic field.
We could also contemplate  extracting from the 2-point function information  about
possible bound states \cite{Caselle02}.   Finally, we note that the BMW method
paves the way towards  understanding   a variety of situations where the  momentum structure 
plays a crucial role. For instance, a method similar in spirit has been applied
successfully to the determination of the fixed point structure  of the Kardar-Parisi-Zhang equation 
\cite{KPZPRL,Canet11TBP} and to the calculation of the  spectral function in a Bose gas \cite{Dupuis2009}.

\section{Acknowledgements}
We thank N. Dupuis for discussions and  remarks on a first version of the manuscript.

\appendix

\section{Notation and conventions}
\label{Notations}

By taking successive functional derivatives of $\Gamma_k[\phi]$ with respect to $\phi(x)$, and then
letting the field be constant, one gets the $n$-point functions 
\beq
\Gamma_k^{(n)}(x_1,\cdots ,x_n,\phi)\equiv \left. \frac{\delta^n\Gamma_k}{\delta\phi(x_1) \dots
\delta\phi(x_n)}\right|_{\phi(x)\equiv \phi}
\eeq
 in a constant background field $\phi$. Since the background is constant, these 
functions are invariant under translations 
of the coordinates, and it is convenient to factor out of the definition of their 
Fourier transform  the $\delta$-function 
that expresses the conservation of the total momentum. Thus, with the usual  abuse 
of notation,  we define the 
 $n$-point functions $\Gamma_k^{(n)}(p_1,\dots,p_n,\phi)$ as: 
\beq\label{gamman} 
&& \hskip -0.5 cm (2\pi)^d \;\delta^{(d)}\Big(\sum_j p_j\Big)
\;\Gamma_k^{(n)}(p_1,\cdots,p_n,\phi)\equiv\qquad\qquad\nonumber\\ 
&& \hskip -0.5 cm \int d^dx_1\dots d^dx_{n}\; e^{i\sum_{j}
p_jx_j}\Gamma_k^{(n)}(x_1,\cdots,x_n,\phi). \notag 
\eeq 
We use here the convention of incoming momenta, and it is understood that 
in $\Gamma_k^{(n)}(p_1,\dots,p_n,\phi)$ the sum of all 
momenta vanishes, so that $\Gamma_k^{(n)}$ is actually a function of $n-1$
 momentum variables (and of $\phi$).  Notice that we use brackets for functional, e.g. $\Gamma_k[\phi]$,
and parenthesis for functions, e.g. $\Gamma_k^{(n)}(p_1,\cdots ,p_n,\phi)$ when $\phi$ is uniform.
 For the 2-point 
function evaluated in a uniform field configuration, which effectively depends on a single momentum $p$, we  often use 
the simplified notation $\Gamma_k^{(2)}(p,\phi)$ 
in place of  $\Gamma_k^{(2)}(p,-p,\phi)$.

%%%%%%%%%%%%%%%%%%%%%%%%%%%%%%
%%%%%%%%%%%%%%%%%%%%%%%%%%%%%

\section{Extension of BMW}
\label{BMWimproved}

In  the approximation  BMW with  $s=2$, we make the  following substitutions in the r.h.s. of the flow 
equation for $\Gamma_k^{(2)}(p)$:
$\Gamma_k^{(4)}(p,-p,q,-q)\longrightarrow    \Gamma_k^{(4)}(p,-p,0,0)$,
and
$\Gamma_k^{(3)}(p,q,-p-q)\longrightarrow\Gamma_k^{(3)}(p,0,-p)$, that is, we set the loop 
momentum $q$ to zero in the 3 and 4-point functions.    (In
this appendix we  do not indicate explicitly the  dependence on $\phi$
of all  $n$-point functions  in order to  alleviate the  notation.) By
doing  so, one  obtains a  closed  equation for  the 2-point  function
$\Gamma_k^{(2)}(p)$, which is the object calculated with optimum accuracy at the level $s=2$.  
As explained in  the main part of the article, the general strategy to obtain the 3 and 4-point 
functions with comparable accuracy is to 
 consider    higher   orders   ($s>2$)   in  the
approximation scheme.  However, in this appendix we show that one can already improve the accuracy 
of  $\Gamma_k^{(3)}$   and
$\Gamma_k^{(4)}$ simply by  exploiting   the information  available  on
$\Gamma_k^{(2)}(p)$.

Let us consider first the function $\Gamma_k^{(4)}$. We know that, at 
one-loop and in vanishing fields, it has the following structure
\begin{eqnarray}
&&\hspace{-.5cm}\Gamma_k^{(4),1\,{\rm loop}}(p_1,p_2,p_3,p_4)\nonumber\\
&&= f(p_1+p_2)+f(p_1+p_3)+f(p_1+p_4),\nonumber\\
\end{eqnarray}
where the function $f(p)$ is easily found to be
\beq
&&f(p)= \frac{1}{2} \Gamma_k^{(4)}(p,-p,0,0)-\frac{1}{6}\Gamma_k^{(4)}(0,0,0,0).\nonumber\\
\eeq
Since
$\Gamma_k^{(4)}(p,-p,0,0)=\del_{\phi}^2 \Gamma_k^{(2)}(p)$ (for constant field $\phi$), we arrive at
the following expression for the 4-point function in terms of the 2-point function $ \Gamma_k^{(2)}(p)$:
\beq
\label{generalGamma4}
&&\Gamma_k^{(4)}(p_1,p_2,p_3,p_4)\approx\frac{1}{2}\del_{\phi}^2\Gamma_k^{(2)}(p_1+p_2)\nonumber\\
&&+\frac{1}{2}\del_{\phi}^2\Gamma_k^{(2)}(p_1+p_3)+\frac{1}{2}\del_{\phi}^2\Gamma_k^{(2)}(p_1+p_4)\nonumber\\
&&-\frac{1}{2}\del_{\phi}^2\Gamma_k^{(2)}(0).
\eeq
Note that this expression is, by
construction, symmetric under the exchange of the external legs, and it is 1-loop exact  at zero external field.  

For the  function $\Gamma_k^{(3)}$, one can extract the following equivalent in the  limit of vanishing field:
\beq
\frac{\Gamma_k^{(3)}(p,q,l)}{\phi} 
&&\sim  \partial_\phi\Gamma_k^{(3)}(p,q,l)\Big|_{\phi=0} \nonumber\\
&&\sim \Gamma_k^{(4)}(p,q,l,0)\Big|_{\phi=0}\ .
\eeq
Then, by using the approximation above for $\Gamma_k^{(4)}$  (\ref{generalGamma4})
one obtains the following expression for $\Gamma_k^{(3)}$, whose zero field equivalent is 
exact at one-loop:
\beq
\label{generalGamma3}
&&\Gamma_k^{(3)}(p,q,l)\approx
\frac{1}{2}\del_\phi\Gamma_k^{(2)}(p)+\frac{1}{2}\del_\phi\Gamma_k^{(2)}(q)\nonumber\\
&&+\frac{1}{2}\del_\phi\Gamma_k^{(2)}(l)
-\frac{1}{2}\del_\phi\Gamma_k^{(2)}(0).
\eeq

At this  point, we note  that we may  use the new expressions  that we
have obtained  for $\Gamma_k^{(3)}$  and $\Gamma_k^{(4)}$ in  the flow
equation for $\Gamma_k^{(2)}$. Since these $n$-point functions are now
one-loop exact, the  resulting approximation for $\Gamma_k^{(2)}$ will
be  2-loop exact in zero external field.  This yields  therefore  an improvement  of the  BMW
approximation, in particular in the high momentum region where we know
that it loses accuracy.

Consider then Eq.~(\ref{gamma2champnonnul}) for $\Gamma_k^{(2)}$, and re-write it in terms of
$\Delta_k(p)$:
\begin{eqnarray}
\label{gamma2champnonnula}
&&\hspace{-.5cm}\partial_t\Delta_{k}(p;\rho)=\int_q\,\partial_t R_k(q)\,G_{k}^2(q)\nonumber\\
&&\times\Big\{[\Gamma_{k}^{(3)}(p,q,-p-q)]^2 G_{k}(q+p) \nonumber\\
&&\hspace{.5cm}-[\Gamma_{k}^{(3)}(0,q,-q)]^2 G_{k}(q)\nonumber\\
&-& \frac{1}{2}[\Gamma_{k}^{(4)}(p,-p,q,-q)-\Gamma_{k}^{(4)}(0,0,q,-q)]\Big\}.\nonumber\\
\end{eqnarray}
Next, perform  the substitutions 
(\ref{generalGamma3}) and \beq
\label{particulargamma4}
\Gamma_k^{(4)}(p,-p,q,-q)&\to&\frac{1}{2}\del_{\phi}^2\Gamma_k^{(2)}(p+q,-p-q)\nonumber\\
&&\hspace{-1cm}+\frac{1}{2}\del_{\phi}^2\Gamma_k^{(2)}(p-q,-p+q).    
\eeq
One then gets
\beq
\partial_t\Delta_{k}(p)= 2\rho H(p)-\frac{1}{2}
L(p),
\eeq
with
\beq
&& H(p)\equiv \int_q\,\partial_t R_k(q)\,G_{k}^2(q)\Big\{ G_{k}(q+p) \big[\frac{1}{2}\Delta_k^\prime(p)\nonumber\\
&&\ \ \ \ \ 
+\frac{1}{2}\Delta_k^\prime(q)+\frac{1}{2}\Delta_k^\prime(p+q)+3 V_k''+2\rho V_k''' \big]^2\nonumber\\
&&\ \ \ \ \ -G_{k}(q)\big[\Delta_k^\prime(q)+3 V_k''+2\rho V_k'''\big]^2
\Big\},
\eeq
and
\beq
L(p)&=&\int_q \, \partial_t R_k(q)\,G_{k}^2(q)
\Big\{ \Delta_k^\prime(p+q)+\nonumber\\
&&\hskip -1cm 2\rho\,\Delta_k^{\prime\prime}(p+q)- \Delta_k^\prime(q)-2\rho\,\Delta_k^{\prime\prime}(q)\Big\}.
 \eeq
It  is  not  difficult  to  generalize  these  expressions to the $O(N)$ model with arbitrary $N$. 
However, we do not present these here because, in spite 
of the good  properties presented above, this extended  version of the
BMW    approximation     proves    to   be   numerically unstable and  we have not been  able to solve
the corresponding equations  with  simple  techniques.  A  further analysis,  using more  
elaborate numerical  techniques, is called for.

%%%%%%%%%%%%%%%%%%%%%%%%%%%%%%%%%
%%%%%%%%%%%%%%%%%%%%%%%%%%%%%%%%%
\section{Integrals}
\label{Integrals}

In this appendix, we give details on the calculation of the  integrals $I_n(k;\rho)$ and  $J_n(p;k;\rho)$.

In the case of the integral $I_n(k;\rho)$, 
since $G_k(q)$ (in a uniform external field) depends only on $q^2$,
the angular integral is straightforward. One gets
\beq
\label{defIa}
&&I_n= \frac{S_d}{(2\pi)^d}\int_0^\infty dq
\,q^{d-1}\,\,\partial_t R_k(q)\,\, G^n_k(q;\rho),\nonumber\\
\eeq
where
\beq
S_d=\frac{2\pi^{d/2}}{\Gamma(d/2)},\qquad K_d=\frac{S_d}{d(2\pi)^d}.
\eeq
In the case of the integral $J_n(p;k;\rho)$, the presence of the external 
momentum $p$ makes the angular integral  more involved. 

\subsection{Angular integrations}

Consider integrals generically of the form
 \beq\label{genericint}
 \int_q \, g(q)\, F(|p+ q|)  \equiv
{\cal I}(p).
\eeq
One can proceed as follows
\beq\label{generic}
 {\cal I}(p)&=&\int_q \, g(q)\,  F(|p+ q|)\nonumber\\
&=&\int_0^\infty dq\,q^{d-1}\, g(q)\,\int \frac{d\Omega_d}{(2\pi)^d} F(|p+ q|) \nonumber\\
&=&\frac{S_{d-1}}{(2\pi)^d}\int_0^\infty  dq\,q^{d-1}\, g(q)\nonumber\\
&&\hskip -0.4cm\times \int_0^\pi d\theta \sin^{d-2}\theta\,
\, F(\sqrt{p^2+q^2+2pq\cos\theta})  \nonumber\\
&=& \frac{S_{d-1}}{(2\pi)^d}\int_0^\infty dq \,q^{d-2}\,
\frac{g(q)}{p}\nonumber\\
&&\times \int_{|p-q|}^{p+q}d\xi \,\xi \,{\cal J}_d(\xi,p,q)\, F(\xi),
\eeq
where we made the change of variables $\xi\equiv \sqrt{p^2+q^2+2pq\cos\theta}$, and
\beq\label{Jacobian}
{\cal J}_d(\xi,p,q)\equiv  \left[
1-\left(\frac{\xi^2-p^2-q^2}{2pq}\right)^2\right]^{\frac{d-3}{2}}.
\eeq
The interest of the last formula (\ref{generic}) lies in the fact 
that the needed integration points belong to the grid so that the 
integral can be calculated numerically without the need of interpolation. 
Furthermore, this method is particularly convenient in $d=3$
 because  the Jacobian (\ref{Jacobian}) is then trivial.

\subsection{Dimensions less than- 3}

The Jacobian (\ref{Jacobian}), is unity in $d=3$, and regular for $d>3$, 
but becomes singular for $d<3$. More
precisely, for $d<3$ it diverges when $\xi$ approaches the
boundaries of its integration domain ($\xi=p+q$ or $|p-q|$). Even if the
integral eventually converges, this divergence is the
source of numerical difficulties.

We then use for $d<3$ a different strategy, based on 
Cartesian variables. We define $q_1$ as the component of $q$ along $p$, and proceed as follows
\beq
 {\cal I}(p)&=&\int\frac{d^{d-1}q_2}{(2\pi)^{d-1}} 
\int_{-\infty}^{+\infty}\frac{dq_1}{2\pi} \, g(q)\,F(|p+q|)\nonumber\\
&=&\frac{S_{d-1}}{(2\pi)^d}\int_0^\infty q_2^{d-2}dq_2\nonumber\\
&&\hskip -0.8cm\times\int_{-\infty}^{\infty} dq_1  \, g(q)\, F\Big(\sqrt{p^2+q_1^2+q_2^2+2pq_1}\Big),\nonumber\\
\eeq
with $q$ the modulus of the vector ${\bf q}$: $q=\sqrt{q_1^2+q_2^2}$ and $|p+q|=\sqrt{p^2+q_1^2+q_2^2+2pq_1}$. 

This expression has no singularities for $d\ge 2$ but it requires 
multiple interpolations that make the numerics
more involved than in $d\ge 3$.

\subsection{Small momentum limits}

The integral $J_n(p;k;\rho)$ is regular when $p\to 0$. However, the expression given by the angular
integration does not make this manifest. 
To get  the small $p$ behavior of the generic integral (\ref{genericint}), one can expand directly
$F(| {\bf p+ q}|)-F(q)$ in the first line of eq.~(\ref{generic}):
\beq
F(| {\bf p+ q}|)&-&F(q)=\frac{2 {\bf p\cdot   q}+  p^2}{2q}\del_q F(q)\nonumber\\
&+&\frac{(  {\bf p}\cdot   {\bf q})^2}{2q^2}\left(\del_q^2
F(q)-\frac{1}{q}\del_q F(q)\right)\nonumber\\
&&+{\cal O}( p^3).
\eeq
Then one can
use, with the brackets denoting angular averages,
\beq
\langle f(q)\rangle&=&\frac{1}{S_d}\int d\Omega_d \,f(q)= f(q) \nonumber\\
\langle ({\bf  q\cdot p})\,f(q)\rangle &=&\frac{1}{S_d}\int d\Omega_d \,({\bf  q\cdot p})\,f(q)=0
\nonumber\\
\langle ({\bf  q\cdot p})^2\,f(q)\rangle &=&\frac{1}{S_d}\int d\Omega_d \,({\bf  q\cdot p})^2\,f(q)
\nonumber\\
&&=\frac{ p^2\,q^{2}}{d} f(q),
\eeq
to obtain
\beq
\label{angularaverage1}
&&\hspace{-.5cm}\left\langle  F(|{\bf p+ q}|)-F(q) \right\rangle\nonumber\\
&&= \frac{p^2}{2d} \left(\del_q^2 F(q)+\frac{d-1}{q}\del_q F(q)\right)+{\cal O}( p^4), \nonumber\\
\eeq
and
\beq\label{genericintegral}
{\cal I}(p)&-&{\cal I}(0)\nonumber\\
&\ \ \ =&\frac{p^2}{2}\,K_d\int_0^\infty dq \,g(q)\,
\,q^{d-1}\,\nonumber\\
&&\times\Big(\del_q^2 F(q)+\frac{d-1}{q}\del_q F(q)\Big)+{\cal O}(p^4)
\nonumber\\
&\ \ \ =&\frac{p^2}{2}\,K_d\int_0^\infty dq \,g(q)\,
\del_q \Big(q^{d-1}\,\del_q  F(q)\Big)\nonumber\\
&&+{\cal O}(p^4).
\eeq

%%%%%%%%%%%%%%%%%%%%%%%%%%%%%%%%%
%%%%%%%%%%%%%%%%%%%%%%%%%%%%%%%%%
\section{Generalization to $O(N)$ models}
\label{ONmodels}

In this appendix the $s=2$ BMW approximation and the corresponding flow equations are presented for $O(N)$ models. 
 The exact flow of the 2-point function in a constant external field  reads
(we omit the renormalization group parameter $k$ in this appendix for notational simplicity):
\begin{eqnarray}
\label{gamma2}
&&\hspace{-.5cm}\partial_t \Gamma_{ab}^{(2)}(p,\bfphi)=\int_q\partial_t (R(q))_{in}
\Big\{G_{ij}(q,\bfphi)\nonumber \\
&&\times \Gamma_{ajh}^{(3)}(p,q,-p-q,\bfphi) G_{hl}(q+p,\bfphi)\nonumber\\
&&\times \Gamma_{blm}^{(3)}(-p,p+q,-q,\bfphi)G_{mn}(q,\bfphi) \nonumber \\
&&-\frac{1}{2}G_{ij}(q,\bfphi)\Gamma_{abjh}^{(4)}(p,-p,q,-q,\bfphi)G_{hn}(q,\bfphi)\Big\},\nonumber\\
\end{eqnarray}
where  $a,b,\ldots$ denote $O(N)$ indices and $\bfphi$  a $N$-component uniform field. 
Within the BMW approximation, we make the substitutions:
\begin{eqnarray}
\Gamma_{ajh}^{(3)}(p,q,-p-q,\bfphi)&&\to
\frac{\partial \Gamma_{ah}^{(2)}(p,-p,\bfphi)} {\partial \phi_{j}},  \nonumber \\
\Gamma_{abjh}^{(4)}(p,-p,q,-q,\bfphi)&&\to
\frac{\partial^2
\Gamma_{ab}^{(2)}(p,-p,\bfphi)} {\partial \phi_j \partial \phi_h}.\nonumber\\
\end{eqnarray}
In order to manifestly preserve the $O(N)$ symmetry along the flow,  
the regulator $\Delta S_k$
has to be an $O(N)$ scalar and, accordingly, the cut-off function a tensor
\[
(R(q))_{ij} \equiv R(q) \delta_{ij}.
\]
The symmetry of the theory also implies that the matrix of 2-point functions can be written in terms
of two independent tensors. We chose to write it in the form
\begin{equation}\label{gamma2_decomp}
\Gamma_{ab}^{(2)} (p, -p,\bfphi) = \Gamma_{A} (p,\rho)\delta_{ab} + \phi_a \phi_b \Gamma_B (p, \rho),
\end{equation}
with $\rho=\frac 1 2\sum_a \phi_a \phi_a$. This form turns out to be convenient in the limit $\rho\to 0$.

The symmetry also allows us to write the propagator in this equation
in terms of its longitudinal and transverse components
with respect to the external field
\beq
G_{ab}(p^2,\bfphi)&=&G_T(p^2,\rho)\left(\delta_{ab}-\frac{\phi_a\phi_b}{2\rho}\right)\nonumber\\
& &+ G_L(p^2,\rho)\frac{\phi_a\phi_b}{2\rho}.
\eeq
\noindent It is easy to find the relationship between these propagators and $\Gamma_A$ and $\Gamma_B$
\begin{align}
G_T^{-1}(p,\rho) = &
\Gamma_A(p,\rho) + R(p), \label{propGTgamma} \\
G_L^{-1}(p,\rho) = &\Gamma_A(p,\rho)
+ 2\rho \Gamma_B(p,\rho)+ R(p).
\label{propGLgamma}
\end{align}
Using the definition (\ref{gamma2_decomp}) of the functions
$\Gamma_A$ and $\Gamma_B$, as well as the form given above for the
propagators, one can decompose the flow equation (\ref{gamma2}) in two equations
for $\Gamma_A$  and $\Gamma_B$.

As in the case $N=1$, we introduce the functions
\begin{align}
\Delta_A(p, \rho) = & \Gamma_A(p, \rho) - p^2  - \Gamma_A(p = 0,
\rho) \label{defdelta_a_2}, \\
\Delta_B(p, \rho) = & \Gamma_B(p,
\rho) - \Gamma_B(p = 0, \rho) \label{defdelta_b_2}.
\end{align}
Notice that at bare level $\Gamma_A(p, \rho)-\Gamma_A(p=0, \rho)=p^2$ while $\Gamma_B(p, \rho)-\Gamma_B(p=0, \rho)=0$, which explains the difference between the two definitions. In terms of these functions, $\Gamma_A$ and $\Gamma_B$ read
\begin{align}
\Gamma_A(p, \rho) = & p^2 + \Delta_A(p, \rho) + V' \label{defdelta_a}, \\
\Gamma_B(p, \rho) = &\Delta_B(p, \rho) + V'',
\label{defdelta_b}
\end{align}
where the primes denote derivatives with respect to $\rho$.
The equations for $\Delta_A$ and $\Delta_B$ read:
\begin{align}\label{flowdeltaA}
&\partial_t \Delta_A(p, \rho) = 2  \rho \Big\{ J_3^{LT} (\Delta_A'+V'')^2 \nonumber\\
&+ J_3^{TL} (\Delta_B+ V'')^2 - (I_3^{LT} +I_{3}^{TL})V''^2 \Big\} \notag \\
&- \frac{1}{2} I_2^{LL} (\Delta_A' + 2 \rho \Delta_A'') \notag \\ &- \frac{1}{2} I_2^{TT} ((N-1) \Delta_A' + 2 \Delta_B),
\end{align}
\begin{align}\label{flowdeltaB}
\partial_t \Delta_B(p, \rho) &= J_3^{TT} (N-1) (\Delta_B + V'')^2  \nonumber\\
&\hskip -0.6cm- J_3^{LT} (\Delta_A' + V'')^2 - J_3^{TL} (\Delta_B+ V'')^2 \notag \\
&\hskip -0.6cm+ J_3^{LL} \Big\{(\Delta_A'+2\Delta_B+3V'')^2 \nonumber\\
&\hskip -0.6cm+ 4 \rho \big(\Delta_B' + V'''\big)\big(\Delta_A'+2\Delta_B+3V''\big) \notag \\
&\hskip -0.6cm+ 4 \rho^2 (\Delta_B' + V''')^2\Big\} - \frac{1}{2} I_2^{TT} (N-1) \Delta_B' \nonumber\\
&\hskip -0.6cm - \frac{1}{2} I^{LL}_2 (5 \Delta_B' + 2 \rho \Delta_B'') \notag \\
&\hskip -0.6cm- \left((N-1) I^{TT}_3-I^{LT}_3-I^{TL}_3\right) V''^2\nonumber \\
&\hskip -0.6cm- I^{LL}_3 (3V''+2\rho V''')^2 + \Delta_B I_{A}, 
\end{align}
where we have omitted the $\rho$ and $p$ dependences on the right hand side for compactness. We have introduced the integrals ($n>1$)
\beq
\label{IJON}
 J_n^{\alpha \beta}(p,\rho) &= & \int_q \partial_t R(q) G_{\alpha}^{n-1}(q,\rho) G_{\beta}(p+q,\rho) , \nonumber\\
I_n^{\alpha \beta}(\rho)&=&J_n^{\alpha \beta}(p=0,\rho), 
\eeq
 with $\alpha$, $\beta$ standing either for $L$ (longitudinal) or $T$ (transversal).  For $n=1$ we set
 \beq
 I_1=(N-1)I_1^{TT}(\rho)+I_1^{LL}(\rho).
 \eeq
It turns out to be useful to also introduce the integral
\begin{multline}
I_{A}(\rho)\equiv\int_q \partial_t R(q) (G_L(q,\rho)+G_T(q,\rho)) \\ \times G_L(q,\rho)G_T(q,\rho),\nonumber\\
\end{multline}
and, in intermediate steps, we have used the identity
\begin{multline}
\frac{1}{\rho} \big(G_T^2(q,\rho) - G_L^2(q,\rho)\big)  = 2G_L(q,\rho) G_T(q,\rho)
 \\
\times \Gamma_B(q,\rho) \big( G_L(q,\rho) + G_T(q,\rho) \big),
\end{multline}
which allows us to handle expressions that are manifestly regular for $\rho = 0$.

As said in the main text, an accurate study of the critical regime requires to use
 dimensionless variables. Using again $W(\rho) = V'(\rho)$  we define:
\begin{align}
 &k^2Z_k\left(\tilde p^2+\tilde{\Delta}_A(\tilde p,\tilde \rho)\right) =  p^2+\Delta_A(p,\rho), \\
&\tilde{\Delta}_B(\tilde p, \tilde \rho) = \frac{K_d\Delta_B(p,\rho)}{Z_k^2 k^{4-d}}.
\end{align}

We also have to use the dimensionless functions corresponding to (\ref{IJON}):
\begin{align}
\tilde{I}_{3}^{\alpha \beta}(\tilde \rho) &= I_{3}^{\alpha \beta}(\rho) \frac{Z_k^2 k^{4-d}}{K_d}, \notag\\
\tilde{J}_{3}^{\alpha \beta} (\tilde p,\tilde \rho)&= J_{3}^{\alpha \beta}(p,\rho) \frac{Z_k^2 k^{4-d}}{K_d},  \\
\tilde{I}_{2}^{\alpha \beta} (\tilde \rho) &= I^{\alpha \beta}_2(\rho) \frac{Z_k k^{2-d}}{K_d}. \notag
\end{align}

For numerical reasons, as explained in the main text for the $N=1$ case, we study the flow of
\begin{align}
\tilde{Y}_A(\tilde p, \tilde \rho) &= \frac{\tilde{\Delta}_A}{\tilde{p}^2}\ , & \tilde{Y}_B(\tilde p, \tilde \rho) 
&= \frac{\tilde{\Delta}_B}{\tilde{p}^2}.
\end{align}

The dimensionless flow equations can then be calculated from equations (\ref{flowdeltaA}) and (\ref{flowdeltaB}):
\begin{widetext}
\begin{align}
\partial_t \tilde{Y}_A(\tilde p,\tilde \rho) &= \eta (\tilde{Y}_A +1)+ \tilde p \frac{\partial \tilde{Y}_A}{\partial \tilde p} + (d-2+\eta)\tilde \rho \tilde{Y}_A'\notag \\ &+ 2  \tilde \rho \Big\{ \tilde{J}_{3}^{LT} \tilde{p}^2 \big(\tilde{Y}_A' + \frac{\tilde{W}'}{\tilde{p}^2}\big)^2 + \tilde{J}_{3}^{TL} \tilde{p}^2 \big(\tilde{Y}_B + \frac{\tilde{W}'}{\tilde{p}^2}\big)^2 - (\tilde{I}_{3}^{LT}+\tilde{I}_{3}^{TL}) \frac{\tilde{W}'^2}{\tilde{p}^2} \Big\} \notag \\ 
&- \frac{1}{2} \tilde{I}_{2}^{LL} (\tilde{Y}_A' + 2 \tilde \rho \tilde{Y}_A'') - \frac{1}{2} \tilde{I}_{2}^{TT} \big((N-1) \tilde{Y}_A' + 2 \tilde{Y}_B\big),
\end{align}
\begin{align}
\partial_t \tilde{Y}_B(\tilde p,\tilde \rho) &= (d -2 + 2\eta) \tilde{Y}_B + \tilde p \frac{\partial \tilde{Y}_B}{\partial \tilde p} + (d-2+\eta)\tilde \rho \tilde{Y}_B'  \notag \\
&+ (N-1) \tilde{J}_{3}^{TT} \tilde{p}^2 \big(\tilde{Y}_B+ \frac{\tilde{W}'}{\tilde{p}^2}\big)^2 + \tilde{J}_{3}^{LL} \Big\{\tilde{p}^2 \big(\tilde{Y}_A'+ 2 \tilde{Y}_B + \frac{3\tilde{W}'}{\tilde{p}^2} \big)^2  \notag \\ &+4 \tilde \rho \tilde{p}^2\big( \tilde{Y}_A' +2 \tilde{Y}_B' \frac{3\tilde{W}}{\tilde{p}^2}\big) \big(\tilde{Y}_B'+\frac{\tilde{W}''}{\tilde{p}^2}  \big) + 4 \tilde{\rho}^2 \tilde{p}^2 \big(\tilde{Y}_B'+\frac{\tilde{W}''}{\tilde{p}^2}  \big)^2 \Big\} \notag \\
& - \tilde{J}_{3}^{LT} \tilde{p}^2 \big(\tilde{Y}_A'+ \frac{\tilde{W}'}{\tilde{p}^2}\big)^2 - \tilde{J}_{3}^{TL} (\tilde{p}^2 \tilde{Y}_B^2 + 2 \tilde{Y}_B \tilde{W}' + \frac{\tilde{W}'^2}{\tilde{p}^2}) \notag \\ &- \frac{\tilde{I}_{3}^{LL}}{\tilde{p}^2} \big(3 \tilde{W}' + 2 \tilde{\rho} \tilde{W}''\big)^2 -\big( (N-1) \tilde{I}_{3}^{TT} -\tilde{I}_{3}^{LT} -\tilde{I}_{3}^{TL}\big)\frac{\tilde{W}'^2}{\tilde{p}^2} \notag \\
&- \frac{1}{2} \tilde{I}_{2}^{TT} (N-1) \tilde{Y}_B' - \frac{1}{2} \tilde{I}_{2}^{LL} (5 \tilde{Y}_B' + 2 \tilde \rho \tilde{Y}_B'') +  \tilde{Y}_B I_A,
\end{align}
\end{widetext}
with the primes now denoting derivatives w.r.t. $\tilde \rho$, and we have omitted the $\tilde \rho$ and $\tilde p$ dependences on the right hand side for compactness.

The flow equation for the potential, which reads
\begin{equation}
\partial_t V(\rho) = \frac{1}{2}I_{1}(\rho),
\end{equation}
allows us to derive an equation for the dimensionless derivative of the potential
\begin{equation}
\partial_t \tilde{W}(\tilde \rho) = -(2-\eta)\tilde{W} + (d-2+\eta)\tilde \rho \tilde{W}' + \frac{1}{2} \frac{\partial \tilde{I}_{1}(\tilde \rho)}{\partial \tilde \rho}.
\end{equation}

The flow of $\eta_k$ follows from fixing a renormalization condition analogous to Eq. (\ref{renorm3}). 
We impose for all values of $k$,
\begin{equation}
\tilde{Y}_A(\tilde p_0,\tilde{\rho}_0) = 0.
\end{equation}
The simplest choice is $\tilde p_0 = 0$ and $\tilde{\rho}_0 = 0$. It leads to
\begin{equation}\label{eq_for_eta}
 \eta_k = \frac{1}{2}(N\tilde{Y}_A' (0,0)+ 2 \tilde{Y}_B(0,0)) \tilde{I}_{2}^{TT}(\tilde{\rho} = 0),
\end{equation}
where we have used $\tilde{I}_2^{TT}(\tilde\rho = 0) = \tilde{I}_2^{LL}(\tilde\rho = 0)$.

In the case of a generic renormalization point, the equation for $\eta_k$ is more cumbersome
\begin{align}
 \eta_k &= \frac{-1}{1+\tilde{\rho}_0 \tilde{Y}_A'} \times \bigg\{ \tilde p_0 \frac{\partial \tilde{Y}_A}{\partial \tilde p} + (d-2)\tilde{\rho}_0 \tilde{Y}_A' \notag \\ 
&\hskip -0.6cm+ 2  \tilde{\rho}_0 \left[ \tilde{J}_{3}^{LT} (\tilde{p_0}^2 \tilde{Y}_A'^2 + 2 \tilde{Y}_A' \tilde{W}' + \frac{\tilde{W}'^2}{\tilde{p_0}^2}) - \tilde{I}_{3}^{LT} \frac{\tilde{W}'^2}{ \tilde{p_0}^2} \right] \nonumber \\
&\hskip -0.6cm+ 2 \tilde{\rho}_0 \left[\tilde{J}_{3}^{TL} (\tilde{p_0}^2 \tilde{Y}_B^2 + 2 \tilde{Y}_B \tilde{W}' + \frac{\tilde{W}'^2}{\tilde{p_0}^2}) - \tilde{I}_{3}^{TL} \frac{\tilde{W}'^2}{\tilde{p_0}^2} \right] \nonumber \\
&\hskip -0.6cm- \frac{1}{2} \tilde{I}_{2}^{LL} (\tilde{Y}_A' + 2 \tilde{\rho}_0 \tilde{Y}_A'') - \frac{1}{2} \tilde{I}_{2}^{TT} ((N-1) \tilde{Y}_A' + 2 \tilde{Y}_B) \bigg\},
\end{align}
with all functions evaluated at $\tilde p=\tilde p_0$, $\tilde{\rho} = \tilde{\rho}_0$.

We also define
\begin{equation}
r_t(\tilde q) = -\eta\tilde q^2 r(\tilde q) - \tilde q^3\partial_{\tilde q} r(\tilde q),
\end{equation}
and the dimensionless propagators
\begin{align}
 \tilde{G}_T(\tilde p, \tilde \rho) &= \frac{1}{\tilde p^2(\tilde{Y}_A+1+ r(\tilde p)) + \tilde{W} }, \label{GTadim} \\
 \tilde{G}_L(\tilde p, \tilde \rho) &= \frac{1}{\tilde p^2(\tilde{Y}_A +1+ 2 \tilde \rho \tilde{Y}_B + r(\tilde p)) + \tilde{W} + 2 \tilde \rho \tilde{W}'}, \label{GLadim}
\end{align}
from which follow the expressions
\begin{align}\label{In_adim}
\tilde{I}_{n}^{\alpha \beta} (\tilde \rho)&= d \int_0^{\infty} d\tilde q\, \tilde q^{d-1}
r_t(\tilde q) \tilde{G}_{\alpha}^{n-1}(\tilde q) \tilde{G}_{\beta}(\tilde q),
\end{align}
\begin{align}
\tilde{J}_{3}^{\alpha \beta}(\tilde p,\tilde \rho)  =& \frac{S_{d-1}}{K_d(2\pi)^d}\int_0^{\infty} d\tilde q\,
\frac{\tilde q^{d-2}}{\tilde p} r_t(\tilde q) \tilde{G}_{\alpha}^2(\tilde q) \notag \\ &\times \int_{|\tilde p-\tilde q|}^{\tilde p+\tilde q} d\xi \, \xi
\mathcal{J}_d(\xi,\tilde p,\tilde q) \tilde{G}_\beta(\xi),
\end{align}
with $\mathcal{J}_d(\xi,\tilde p,\tilde q)$ as defined in Eq. (\ref{Jacobian}). We also need the functions
\begin{align}
\tilde I_A (\tilde \rho)=& \int_0^{\infty} d \tilde q \; \tilde q^{d-1}
\Big\{r_t(\tilde q) (\tilde G_L(\tilde q) \notag \\ 
&+ \tilde G_T(\tilde q))
\tilde G_L(\tilde q) \tilde G_T(\tilde q) (\tilde{Y}_B(\tilde q)\tilde q^2 + \tilde{W}') \Big\},
\end{align}
\begin{multline}
\tilde K^{\alpha \beta} (\tilde \rho)= \frac{1}{2dK_d}\frac{S_d}{(2\pi)^d} \int_0^\infty
d\tilde q \; r_t(\tilde q) \tilde G_\alpha(\tilde q)  \\ \times \partial_{\tilde q}\left(\tilde q^{d-1} \partial_{\tilde q} \tilde G_\beta (\tilde q) \right),
\end{multline}
that are used in the small momentum region  of the flow equations (cf. Appendix \ref{Integrals} in the $N=1$ case).

%%%%%%%%%%%%%%%%%%%%%%%%%%%%%%%%%
%%%%%%%%%%%%%%%%%%%%%%%%%%%%%%%%%

%\bibliographystyle{unsrt}


\begin{thebibliography}{10}

\bibitem{Wetterich93}  C. Wetterich, Phys. Lett., {\bf B301}, 90 (1993).

\bibitem{Ellwanger93}  U. Ellwanger, Z. Phys., {\bf C58}, 619 (1993).

\bibitem{Wilson}
K. Wilson and J. Kogut, Phys. Rep. C 12, 75 (1974); F. J. Wegner
and A. Houghton, Phys. Rev. A 8, 401 (1973); J. Polchinski,
Nucl. Phys. B 231, 269 (1984).

\bibitem{Berges02} J. Berges, N. Tetradis and C. Wetterich,  Phys. Rep. {\bf 363}, 223 (2002).

\bibitem{delamotte03} 
M. Tissier, B. Delamotte and D. Mouhanna, Phys. Rev. Lett. {\bf 84},  5208 (2000);
B.~Delamotte, D.~Mouhanna and M.~Tissier, \newblock{Phys. Rev. B} {\bf 69}, 134413 (2004);
N.~Dupuis,
%\newblock Unified picture of superfluidity: From bogoliubov's approximation to  popov's hydrodynamic theory.
\newblock {Phys. Rev. Lett.}  {\bf 102}, 190401 (2009);
%\newblock Infrared behavior and spectral function of a bose superfluid at zero  temperature.
A.~Ran\c{c}on and N.~Dupuis,
%\newblock Nonperturbative renormalization group approach to the bose-hubbard  model.
\newblock {Phys. Rev. B}  {\bf 83}, 172501 (2011);
 L. Canet, H. Chat\'e and B. Delamotte, Phys. Rev. Lett. {\bf 92},  255703 (2004);
L. Canet, H. Chat\'e, B. Delamotte, I. Dornic and M. A. Mu\~noz, Phys. Rev. Lett. {\bf 95},  100601  (2005).

\bibitem{ref6}
S. Seide and C. Wetterich, Nucl. Phys. B 562, 524 (1999);
L. Canet, H. Chat ́ , and B. Delamotte, Phys. Rev. Lett. 92,
255703 (2004); T. Machado and N. Dupuis, Phys. Rev. E 82,
041128 (2010).


%%%%%%%%%%%%%%%%  ref 7 %%%%%%%%%%%%%%%%%%%%%%%%%%%%%%%
\bibitem{Blaizot:2005xy}
  J-.~P.~Blaizot, R.~Mend\'ez-Galain and N.~Wschebor,
  %``A new method to solve the non perturbative renormalization group
  %equations,''
  Phys.\ Lett.\ B {\bf 632}, 571 (2006).
  %[arXiv:hep-th/0503103].
  %%CITATION = HEP-TH 0503103;%%


%%%%%%%%%%%%%%%%  ref 8 %%%%%%%%%%%%%%%%%%%%%%%%%%%%%%%
  %\cite{Blaizot:2005wd}
\bibitem{Blaizot:2005wd}
  J-.~P.~Blaizot, R.~Mend\'ez-Galain and N.~Wschebor,
  %``Non perturbative renormalisation group and momentum dependence of n-point
  %functions. I,''
  Phys. Rev. E {\bf 74}, 051116 (2006), Ibid. {\bf 74}, 051117 (2006).
  %%CITATION = HEP-TH 0512317;%%





%%%%%%%%%%%%%%%%   %%%%%%%%%%%%%%%%%%%%%%%%%%%%%%%
\bibitem{Parola} A. Parola and L. Reatto, Adv. Phys. {\bf 44}, 211  (1995);
A. Parola, L. Reatto,  and D. Pini, Phys. Rev. E. {\bf 48}, 3321 (1993).


%%%%%%%%%%%%%%%%  ref  %%%%%%%%%%%%%%%%%%%%%%%%%%%%%%%
%Title: Correlation functions in the Non Perturbative Renormalization Group and field expansion
\bibitem{Guerra} D. Guerra, R. Mend\'ez-Galain and N. Wschebor, Eur. Phys. J. B {\bf 59}, 357 (2007).


%%%%%%%%%%%%%%%%  ref  %%%%%%%%%%%%%%%%%%%%%%%%%%%%%%%
\bibitem{BMWnum}
  J-.~P.~Blaizot, R.~M\'endez-Galain and N.~Wschebor,
 Eur. Phys. J. B {\bf 58}, 297 (2007);
 F. Benitez, R. M\'endez-Galain and N. Wschebor,
 Phys. Rev. B {\bf 77}, 024431 (2008).


%%%%%%%%%%%%%%%%  ref  %%%%%%%%%%%%%%%%%%%%%%%%%%%%%%%
%	Title: Solutions of renormalization-group flow equations with full momentum dependence
\bibitem{Benitez09} F. Benitez,  J.-P. Blaizot, H. Chat\'e,  B. Delamotte, R.~M\'endez-Galain and N.~Wschebor,
 Phys. Rev. E {\bf 80}  030103(R) (2009).


%%%%%%%%%%%%%%%%  ref  %%%%%%%%%%%%%%%%%%%%%%%%%%%%%%%

 %\cite{Tetradis:1993ts}
\bibitem{Tetradis94}
  N.~Tetradis and C.~Wetterich,
  %``Critical exponents from effective average action,''
  Nucl.\ Phys.\ B {\bf 422},  541 (1994).
  %[arXiv:hep-ph/9308214].
  %%CITATION = HEP-PH 9308214;%%


\bibitem{Morris94}  T. R. Morris, Int. J. Mod. Phys., {\bf A9}, 2411 (1994).

 \bibitem{Morris94c}
T.~R. Morris,
%\newblock Derivative expansion of the exact renormalization group.
Phys. Lett. B {\bf 329}, 241 (1994).


\bibitem{Bagnuls:2000ae}
C.~Bagnuls and C.~Bervillier,
%``Exact renormalization group equations: An introductory review,''
Phys.\ Rept.\  {\bf 348}, 91 (2001).

\bibitem{delamotte07}
B. Delamotte, arXiv:cond-mat/0702365.


%%%%%%%%%%%%%%%%  ref  %%%%%%%%%%%%%%%%%%%%%%%%%%%%%%%
\bibitem{tarjus04} 
G. Tarjus and M. Tissier, \newblock {Phys. Rev. Lett.}  {\bf 93}, 267008 (2004);
M. Tissier and G. Tarjus,
%\newblock Unified picture of ferromagnetism, quasi-long-range order and  criticality in random-field models.
\newblock { Phys. Rev. Lett.}  {\bf 96}, 087202 (2006);
G. Tarjus and M. Tissier,
%\newblock Nonperturbative functional renormalization group for random field  models and related disordered systems. i. effective average action formalism.
\newblock { Phys. Rev. B}  {\bf 78}, 024203 (2008);
M. Tissier and G. Tarjus,
%\newblock Nonperturbative functional renormalization group for random field models and related disordered systems. ii. results for the random field $ o(n) $ model.
\newblock {Phys. Rev. B}  {\bf 78}, 024204 (2008);
 J.-P. Kownacki and D. Mouhanna, Phys. Rev. E {\bf 79}, 040101 (2009); K.~Essafi, J.-P. Kownacki and D.~Mouhanna,
%\newblock Crumpled-to-tubule transition in anisotropic polymerized membranes: Beyond the $?$ expansion.
\newblock {Phys. Rev. Lett.}  {\bf 106}, 128102 (2011);
L. Canet, B. Delamotte, O. Deloubri\`ere and N. Wschebor, Phys. Rev. Lett. {\bf 92}, 
 195703  (2004); L. Canet, H. Chat\'e, B. Delamotte and N. Wschebor, Phys. Rev. Lett. {\bf 104}, 150601 (2010). 


%%%%%%%%%%%%%%%%  ref  %%%%%%%%%%%%%%%%%%%%%%%%%%%%%%%
\bibitem{Canet03} L. Canet, B. Delamotte, D. Mouhanna and J. Vidal, Phys. Rev. D {\bf 67}, 065004 (2003);
%\newblock Nonperturbative renormalization group approach to the ising model: a
%  derivative expansion at order $\partial^4$.
 Phys. Rev. B {\bf 68}, 064421 (2003); L. Canet, Phys. Rev. B {\bf 71}, 012418 (2005). 

\bibitem{ref20}
 C. Bagnuls and C. Bervillier, Condens. Matter Phys. 3, 559
(2000).


%%%%%%%%%%%%%%%%  ref  %%%%%%%%%%%%%%%%%%%%%%%%%%%%%%%
 \bibitem{Stevenson} P. Stevenson, Phys. Rev. D {\bf 23}, 2916 (1981).
%Title: Optimization of the derivative expansion in the nonperturbative renormalization group


%%%%%%%%%%%%%%%%  ref  %%%%%%%%%%%%%%%%%%%%%%%%%%%%%%%
\bibitem{TBP} L. Canet, H. Chat\'e, B. Delamotte and C. Gombaud, {\it in preparation}.


\bibitem{arisue2004} H. Arisue, T. Fujiwara and K. Tabata, Nucl. Phys. B (Proc. Suppl.) {\bf 129}, 774 (2004).

%%%%%%%%%%%%%%%%%%%% ajout de ref ici%%%%%%%%%%%%%%%%%%%%%%%%%%%%%

\bibitem{D'Attanasio97}
M. D'Attanasio and T.~R. Morris,
%\newblock Large {N} and the renormalization group.
Phys. Lett. B {\bf 409}, 363 (1997).

%%%%%%%%%%%%%%%%  ref  %%%%%%%%%%%%%%%%%%%%%%%%%%%%%%%
\bibitem{ZINN} J. Zinn-Justin, {\it Quantum Field Theory and Critical Phenomena}, 
(Clarendon Press, Oxford, 2002).

%\cite{Moshe:2003xn}
\bibitem{Moshe03}
  M.~Moshe and J.~Zinn-Justin,
  %``Quantum field theory in the large N limit: A review,''
  Phys.\ Rept.\  {\bf 385}, 69 (2003).
 % [arXiv:hep-th/0306133].
  %%CITATION = HEP-TH 0306133;%%

\bibitem{pogorelov07} A.A. Pogorelov and I.M. Suslov, JETP {\bf 105}, 360 (2007).


%%%%%%%%%%%%%%%%  ref  %%%%%%%%%%%%%%%%%%%%%%%%%%%%%%%
%\cite{Von Gersdorff:2000kp}
\bibitem{Gersdorff00}
  G.~Von Gersdorff and C.~Wetterich,
  %``Nonperturbative renormalization flow and essential scaling for the
  %Kosterlitz-Thouless transition,''
  Phys. Rev. B {\bf 64}, 054513 (2001).
%  [arXiv:hep-th/0008114].


\bibitem{Suslov08} A.A. Pogorelov and I.M. Suslov, J. of Experimental and
Theoretical Physics {\bf 106}, 1118 (2008).

\bibitem{grassberger} P. Grassberger, P. Sutter and L. Sch\"afer, J. Phys. A
{\bf 30}, 7039 (1997).

\bibitem{hasenbusch10} M. Hasenbusch, Phys. Rev. B {\bf 82}, 174433  (2010).

\bibitem{Campostrini06}
% The critical exponents of the superfluid transition in He4
M. Campostrini, M. Hasenbusch, A. Pelissetto and E. Vicari, Phys. Rev. B
{\bf 74}, 144506  (2006).

\bibitem{Campostrini01}
% Critical exponents and equation of state of the three-dimensional Heisenberg universality class
M. Campostrini, M. Hasenbusch, A. Pelissetto, P. Rossi and E. Vicari
Phys. Rev. B {\bf 65}, 144520  (2002).

\bibitem{Guida98}
  R.~Guida and J.~Zinn-Justin,
  %``Critical Exponents of the N-vector model,''
  J.\ Phys.\ A  {\bf 31}, 8103  (1998).
  %%CITATION = JPAGB,A31,8103;%%

\bibitem{Hasenbusch01}
% Eliminating leading corrections to scaling in the 3-dimensional O(N)-symmetric phi^4 model: N=3 and 4
M.~ Hasenbusch,
J. Phys. A {\bf 34}, 8221  (2001).

\bibitem{Antonenko98}
  S.~A.~Antonenko and A.~I.~Sokolov,
  %``Critical exponents for 3D O(n)-symmetric model with n > 3,''
  Phys.\ Rev.\  E {\bf 51}, 1894  (1995).
 % [arXiv:hep-th/9803264].
  %%CITATION = PHRVA,E51,1894;%%


\bibitem{Pelisseto07} A. Pelissetto and E. Vicari, J. Phys. A {\bf 40}, F539 (2007).

\bibitem{Blaizot04} J.-P. Blaizot, R. Mend\'ez Galain and N. Wschebor, Europhys. Lett. {\bf 72}, 705 (2005).


\bibitem{Baym99}
  G.~Baym, J-.~P.~Blaizot, M.~Holzmann, F.~Laloe and D.~Vautherin,
  %``The transition temperature of the dilute interacting Bose gas,''
  %%CITATION = COND-MAT/9905430;%%
Phys. Rev. Lett. {\bf 83}, 1703 (1999). 

\bibitem{Baym00} G. Baym, J.-P. Blaizot and J. Zinn-Justin, Europhys.
    Lett. {\bf 49}, 150 (2000).

\bibitem{ref41}
S. Ledowski, N. Hasselmann, P. Kopietz, Phys. Rev. A 69,
061601(R) (2004).


\bibitem{Kastening03}
B.~M.~Kastening,
%``Bose-Einstein Condensation Temperature of Homogenous Weakly Interacting Bose
%Gas in Variational Perturbation Theory Through Seven Loops,''
Phys.\ Rev.\ A {\bf 69}, 043613 (2004).
%%CITATION = COND-MAT 0309060;%%

\bibitem{Sun02} X.~Sun, Phys. Rev. E {\bf 67}, 066702 (2003).

\bibitem{Arnold01} P. Arnold and G. Moore,
Phys. Rev. Lett. {\bf 87}, 120401 (2001).

\bibitem{Kashurnikov01} V.A.  Kashurnikov, N.~V.  Prokof'ev and B.~V.
Svistunov, Phys. Rev. Lett. {\bf 87}, 120402 (2001).







%Critical structure factor in Ising systems
\bibitem{martin-mayor02} V. Martin-Mayor, A. Pelissetto and E. Vicari, Phys. Rev. E {\bf 66} (2002) 026112.

  
\bibitem{fisher68} M.E. Fisher and J.S. Langer, Phys. Rev. Lett. {\bf 20}, 665 (1968).

\bibitem{bray76} A.J. Bray, Phys. Rev. B {\bf 76}, 1248 (1976).

\bibitem{damay98} P. Damay, F. Leclercq, R. Magli, F. Formisano and P. Lindner, Phys. Rev. B {\bf 58}, 12038 (1998).


\bibitem{ferrell75} R.A. Ferrell and D.J. Scalapino, Phys. Rev. Lett. {\bf 34}, 200 (1975).


\bibitem{Campostrini02b} M. Campostrini, A. Pelissetto, P. Rossi and E. Vicari
 Phys. Rev. E {\bf 65}, 066127  (2002).


\bibitem{Dupuis2009} N. Dupuis, Phys. Rev. A  {\bf 80}, 043627 (2009).

%Bound states and glueballs in three-dimensional Ising systems
\bibitem{Caselle02} M. Caselle, M. Hasenbusch, P. Provero and K. Zarembo, Nucl. Phys. B {\bf 623}, 474 (2002). 


\bibitem{KPZPRL} L. Canet, H. Chat\'e, B. Delamotte and N. Wschebor, 
Phys. Rev. Lett. {\bf 104}, 150601 (2010).

\bibitem{Canet11TBP} L. Canet, H. Chat\'e, B. Delamotte and N. Wschebor, arXiv:1107.2289.


\bibitem{Peli-rev} A. Pelissetto and E. Vicari, Phys.\ Rept. {\bf 368}, 549 (2002). 




\end{thebibliography}
\end{document}